%% file: ms_covariance_gridSPT_v2.tex
\documentclass[preprintnumbers, aps,prd,floatfix,groupedaddress,nofootinbib,superscriptaddress,twocolumn,tightenlines,10pt]{revtex4-1}
\usepackage{graphicx,amsmath,amssymb,amsfonts}
\usepackage{aas_macros}
\usepackage{hyperref}
\usepackage{enumerate} % advanced enumerate environment
\usepackage{color} % for color
\usepackage{bm}% bold math
\usepackage[normalem]{ulem}
\newcommand{\bfv}{{\boldsymbol v}}

\newcommand{\bfx}{{\boldsymbol x}}
\newcommand{\bfk}{{\boldsymbol k}}
\newcommand{\bfn}{{\boldsymbol n}}
\newcommand{\bfu}{{\boldsymbol u}}
\newcommand{\gridspt}{{\tt GridSPT}}
\newcommand{\cov}{\mbox{Cov}}
\newcommand{\covSPTL}{\mbox{Cov}^{\rm\small SPT}_{\rm\small Lin}}
\newcommand{\covSPTNLO}{\mbox{Cov}^{\rm\small SPT}_{\rm\small NLO}}
\newcommand{\covSPTNNLO}{\mbox{Cov}^{\rm\small SPT}_{\rm\small NNLO}}

\newcommand{\deltaw}{\delta_{\rm w}}
\newcommand{\dellocal}{\overline{\delta}}

\newcommand{\deltaK}{\delta^{\rm K}}
\begin{document}
\title{Covariance of the matter power spectrum including the survey window function effect: N-body simulations vs. fifth-order perturbation theory on grids}

\author{Atsushi Taruya}
\affiliation{Center for Gravitational Physics, Yukawa Institute for Theoretical Physics, Kyoto University, Kyoto 606-8502, Japan}
\affiliation{Kavli Institute for the Physics and Mathematics of the Universe, Todai Institutes for Advanced Study, the University of Tokyo, Kashiwa, Chiba 277-8583, Japan (Kavli IPMU, WPI)}
\author{Takahiro Nishimichi}
\affiliation{Center for Gravitational Physics, Yukawa Institute for Theoretical Physics, Kyoto University, Kyoto 606-8502, Japan}
\affiliation{Kavli Institute for the Physics and Mathematics of the Universe, Todai Institutes for Advanced Study, the University of Tokyo, Kashiwa, Chiba 277-8583, Japan (Kavli IPMU, WPI)}
\author{Donghui Jeong}
\affiliation{Department of Astronomy and Astrophysics and Institute for Gravitation and the Cosmos, The Pennsylvania State University, University Park, PA 16802, USA}

%%%%%%%%%%%%%%%%%%%%%%%%%%%%%%%%%%%%%%%%%%%%%%%%%%%%%%%%%%%%%%%%%%%%%%%
\date{\today}
\begin{abstract}
We present a Next-to-next-to-leading (fifth or NNLO) order calculation for the covariance matrix of the matter power spectrum, taking into account the effect of survey window functions. Using the grid-based calculation scheme for the standard perturbation theory, \gridspt, we quickly generate multiple realizations of the nonlinear density fields to fifth order in perturbation theory, then estimate the power spectrum and the covariance matrix from the sample. To the end, we have obtained the non-Gaussian covariance originated from the one-loop trispectrum without explicitly computing the trispectrum. By comparing the \gridspt\, calculations with the $N$-body results, we show that NNLO \gridspt\, result reproduces the $N$-body results on quasi-linear scales where SPT accurately models nonlinear matter power spectrum. Incorporating the survey window function effect to \gridspt\, is rather straightforward, and the resulting NNLO covariance matrix also matches well with the $N$-body results.
\end{abstract}
%%%%%%%%%%%%%%%%%%%%%%%%%%%%%%%%%%%%%%%%%%%%%%%%%%%%%%%%%%%%%%%%%%%%%%%

%\pacs{98.80.-k, 98.62.Py, 98.65.-r}
%\keywords{cosmology, large-scale structure}
\preprint{YITP-20-88}
\maketitle

%%%%%%%%%%%%%%%%%%%%%%%%%%%%%%%%%%%%%%%%%%%%%%%%%%%%%%%%%%%%%%%%%%%%%%%
%%%%%%%%%%%%%%%%%%%%%%%%%%%%%%%%%%%%%%%%%%%%%%%%%%%%%%%%%%%%%%%%%%%%%%%
\section{Introduction}
\label{sec:introduction}
%%%%%%%%%%%%%%%%%%%%%%%%%%%%%%%%%%%%%%%%%%%%%%%%%%%%%%%%%%%%%%%%%%%%%%%
%%%%%%%%%%%%%%%%%%%%%%%%%%%%%%%%%%%%%%%%%%%%%%%%%%%%%%%%%%%%%%%%%%%%%%%

The large-scale matter inhomogeneities probed by galaxy redshift surveys offer various opportunities to test and constrain the cosmology through the structure formation of the Universe. Currently, the main targets of the large-scale structure observations are the acoustic signature imprinted on the spatial clustering of galaxies called the baryon acoustic oscillations (BAO) (e.g., Refs.~\cite{Peebles_Yu1970,EisensteinHu1998}), and the clustering anisotropies due to the peculiar-velocity effect called redshift-space distortions (RSD) (e.g., \cite{Kaiser1987,Hamilton_RSD_review1998}). The precision measurements of BAO and RSD have been providing essential clues to clarify the nature of cosmic acceleration and gravity on cosmological scales (Ref.~\cite{Weinberg_etal2013} for a review). In doing so, the statistical analysis using the galaxy power spectrum (the Fourier transform of the galaxy two-point correlation function) plays a key role to quantify the precision and uncertainty of cosmological parameter estimation. Provided the measured power spectrum, the accuracy and precision of cosmological parameter estimation critically depend on the accuracy of the theoretical model template for the power spectrum and its covariance matrix, which characterize the statistical uncertainties. While the former affects the best-fitting values, the latter mainly changes statistical uncertainties and degeneracy structure of the resulting cosmological constraints.

The theoretical modeling of the observed power spectrum and its covariance matrix, in general, requires inputs both from the observational side and the theoretical sides. That is, we have to include the nonlinear growth of the large-scale structure, as well as the survey window function that encodes the details of the surveying conditions. These effects are particularly manifested in the off-diagonal components of the covariance matrix, which are usually zero for the Gaussian random fields without survey window functions. While a common way to estimate the power spectrum covariance is to use a large number of power spectra measured from the cosmological $N$-body simulations (e.g., Refs.~\cite{Takahashi2009,Blot2015}), it is computationally expensive to generate sufficiently many simulations, each of which covers a large cosmological volume to ensure the reliability of the uncertainty estimation \cite{Dodelson/Schneider:2013}. On the other hand, the analytical calculation with perturbation theory (PT) provides a faster way to compute the covariance matrix in the weakly nonlinear regime \cite{Scoccimarro1999}. Although the applicable range of PT is limited in general, techniques to improve the predictions have been proposed, and they succeeded to model and build the covariance on the scales of our interest \cite{Barreira2017,Mohammed_etal2017}. Recently, based on the Feldman-Kaiser-Peacock (FKP) estimator \cite{Feldman_Kaiser_Peacock1994}, Ref.~\cite{Digvijay_Scoccimarro2019} has computed the power spectrum covariance in redshift space, including most of the possible nonlinear systematics at tree level (leading order). Further, Ref.~\cite{Sugiyama_etal2019} has presented the tree-level calculation for covariance matrices of both the power spectrum and bispectrum, taking into account the RSD and the Poisson sampling noise (shot noise). While the analytical calculation of the covariance matrix in PT framework is compelling, beyond the tree-level order, in practice, a rigorous treatment of PT including all possible contributions is still challenging and requires rather cumbersome calculations (see Ref.~\cite{Bertolini2016}).

In this paper, we present an alternative approach of calculating the covariance matrix of the matter power spectrum, taking advantages of both the numerical and analytical treatments. Recently, in Ref.~\cite{Taruya_Nishimichi_Jeong2018}, we have developed a grid-based algorithm for the standard PT (SPT) calculations called \gridspt, which generates a random realization of non-linear density fields at each grid point. Making use of the fast Fourier transform (FFT), the {\tt C++} implementation of the algorithm quickly generates the nonlinear density fields. By using exactly the same initial conditions as used in $N$-body simulations, we have tested the fifth-order {\gridspt} result in its morphology and statistics in comparison with $N$-body simulations and Lagrangian PT predictions \cite{Taruya_Nishimichi_Jeong2018}. We find an excellent agreement between the {\gridspt} result and the full $N$-body simulations in the weakly nonlinear regime.

As a next step toward practical applications of \gridspt\ in the cosmological data analysis, in this paper, we shall present a proof-of-concept study on how the {\tt GridSPT} method is useful to estimate the power spectrum covariance in the presence of survey window function. In particular, we quantitatively discuss how \gridspt~ properly reproduces the mode-coupling structures in the off-diagonal part of the power spectrum covariance arising both from the nonlinear gravitational evolution and survey window function.

The organization of this paper is as follows. In Sec.~\ref{sec:GridSPT}, we begin by briefly reviewing the grid-based SPT calculation of large-scale structure. Then, in Sec.~\ref{sec:GridSPT_covariance}, we consider how the power spectrum covariance can be perturbatively reconstructed from the grid-based SPT calculations, also presenting the relevant trispectrum formulas at one-loop {next-to-leading} order. The implementation of the survey window effect is also discussed. Then, Sec.~\ref{sec:results} presents the results of the explicit demonstration on the covariance estimation with \gridspt, for which we make a detailed comparison with $N$-body simulations. Finally, Sec.~\ref{sec:conclusion} is devoted to the conclusion and discussions.

%%%%%%%%%%%%%%%%%%%%%%%%%%%%%%%%%%%%%%%%%%%%%%%%%%%%%%%%%%%%%%%%%%%%%%%
%%%%%%%%%%%%%%%%%%%%%%%%%%%%%%%%%%%%%%%%%%%%%%%%%%%%%%%%%%%%%%%%%%%%%%%
\section{Grid-based perturbation theory}
\label{sec:GridSPT}
%%%%%%%%%%%%%%%%%%%%%%%%%%%%%%%%%%%%%%%%%%%%%%%%%%%%%%%%%%%%%%%%%%%%%%%
%%%%%%%%%%%%%%%%%%%%%%%%%%%%%%%%%%%%%%%%%%%%%%%%%%%%%%%%%%%%%%%%%%%%%%%

In this section, we present a succinct review on the grid-based calculation for perturbation theory of large-scale structure named \gridspt, described in Ref.~\cite{Taruya_Nishimichi_Jeong2018}. In essence, \gridspt\, enables us to perform SPT calculations at the field-level (at grid points), and provides a way to generate the higher-order density and velocity fields starting with random realizations of the linear Gaussian density field. The heart of the algorithm is the real-space recursion relation in
Eq.~(\ref{eq:recursion_formula}).

The framework of SPT calculations relies on the single-stream treatment of the cosmological Vlasov-Poisson equations as the basic equations describing the gravitational evolution of matter distribution \cite{Bernardeau:2001qr}. With the single-stream treatment, the large-scale matter inhomogeneities in the cold dark matter (CDM) dominated Universe is described by the pressureless fluid equations coupled with the Poisson equation. Under the irrotational flow assumption valid at large scales, the system of equations describing the nonlinear evolution of density and velocity fields is further reduced to 
%%%%%%%%%%%%%%%%%%%%%%%%%%%%%%%%%%%%%%%%%%%%%%%%%%%
\begin{align}
&\frac{d}{d\eta}\left(
\begin{array}{c}
\delta(\bfx)
\\
\\
\theta(\bfx)
\end{array}
\right)+\Omega_{ab}(\eta)\,\left(
\begin{array}{c}
\delta(\bfx)
\\
\\
\theta(\bfx)
\end{array}
\right)
\nonumber
\\
&\qquad\qquad=\left(
\begin{array}{c}
{\displaystyle (\nabla\delta)\cdot\bfu+\delta\,\,\theta}
\\
\\
{\displaystyle (\partial_ju_k)(\partial_ku_j)+(\nabla\theta)\cdot\bfu}
\end{array}
\right),
\label{eq:basic_PT_eqs}
\end{align}
%%%%%%%%%%%%%%%%%%%%%%%%%%%%%%%%%%%%%%%%%%%%%%%%%%%
where we introduce the time variable $\eta$ defined by $\eta\equiv\ln D_+(t)$ with $D_+$ being the linear growth factor. We denote the comoving coordinate as $\bfx$. The quantities $\delta$ and $\theta$ are the mass density and the velocity-divergence fields, respectively, the latter of which is related to the velocity field $\bfv$ through 
$\theta \equiv-\nabla\bfv/(f\,aH)\equiv \nabla\cdot{\bfu}$ with $f$ being the linear growth rate, defined by $f\equiv d\ln\,D_+/d\ln a$. The field $\bfu$ is the {\it reduced} velocity field, and the irrotational flow implies $\bfu=\nabla[\nabla^{-2}\theta]$. In Eq.~(\ref{eq:basic_PT_eqs}), the matrix $\Omega_{ab}$ generally depends on cosmology and time, but as an approximation, one may replace it with the time-independent constant matrix $\Omega_{ab}^{\rm EdS}$ in the Einstein-de Sitter Universe:
%%%%%%%%%%%%%%%%%%%%%%%%%%%%%%%%%%%%%%%%%%%%%%%%%%%
\begin{align}
 \Omega_{ab}^{\rm EdS}=\left(
\begin{array}{cc}
0 & \qquad -1
\\
\\
{\displaystyle -\frac{3}{2}} & \qquad {\displaystyle \frac{1}{2}}
\end{array}
\right).
\label{eq:Omega_ab_EdS}
\end{align}
%%%%%%%%%%%%%%%%%%%%%%%%%%%%%%%%%%%%%%%%%%%%%%%%%%%
This approximation is shown to give a sufficiently accurate perturbative prediction in a wide class of cosmology close to the $\Lambda$CDM model (e.g., \cite{Pietroni:2008jx,Takahashi2008,Hiramatsu_Taruya2009}).  

We obtain the perturbative solutions for Eq.~(\ref{eq:basic_PT_eqs}) by expanding the density and velocity fields. For the dominant growing-mode contributions, we have 
%%%%%%%%%%%%%%%%%%%%%%%%%%%%%%%%%%%%%%%%%%%%%%%%%%%
\begin{align}
&\delta(\bfx) = \sum_n\,e^{n\,\eta}\,\delta_n(\bfx), \qquad
\theta(\bfx) = \sum_n\,e^{n\,\eta}\,\theta_n(\bfx), \nonumber
\\
&\bfu(\bfx) = \sum_n\,e^{n\,\eta}\,\bfu_n(\bfx). 
\label{eq:PT_expansion}
\end{align}
%%%%%%%%%%%%%%%%%%%%%%%%%%%%%%%%%%%%%%%%%%%%%%%%%%%
Substituting Eq.~(\ref{eq:PT_expansion}) into Eq.~(\ref{eq:basic_PT_eqs}) with Eq.~(\ref{eq:Omega_ab_EdS}), the order-by-order calculation leads to the following recursion relation \cite{Taruya_Nishimichi_Jeong2018}: 
%%%%%%%%%%%%%%%%%%%%%%%%%%%%%%%%%%%%%%%%%%%%%%%%%%%
\begin{align}
&\left(
\begin{array}{c}
{\displaystyle \delta_n(\bfx) }
\\
\\
{\displaystyle \theta_n(\bfx) }
\end{array}
\right) = \frac{2}{(2n+3)(n-1)}\,\left(
\begin{array}{cc}
{\displaystyle n+\frac{1}{2}} & \qquad 1
\\
\\
{\displaystyle \frac{3}{2} } & \qquad n
\end{array}
\right) 
\nonumber
\\
&\times\,\sum_{m=1}^{n-1} \left(
\begin{array}{c}
{\displaystyle (\nabla\delta_m)\cdot\bfu_{n-m} + \delta_m\,\,\theta_{n-m}}
\\
\\
{\displaystyle [\partial_j(\bfu_{m})_k][\partial_k(\bfu_{n-m})_j]+\bfu_m\cdot(\nabla\theta_{n-m}) }
\end{array}
\right),
\label{eq:recursion_formula}
\end{align}
%%%%%%%%%%%%%%%%%%%%%%%%%%%%%%%%%%%%%%%%%%%%%%%%%%%
for $n\geq2$. For the linear-order quantities ($n=1$), the growing-mode initial condition implies
%%%%%%%%%%%%%%%%%%%%%%%%%%%%%%%%%%%%%%%%%%%%%%%%%%%
\begin{align}
\left(
\begin{array}{c}
\delta_1(\bfx)
\\
\\
\theta_1(\bfx)
\end{array}
\right) = \left(
\begin{array}{c}
1
\\
\\
1
\end{array}
\right) \delta_0(\bfx),
\label{eq:recursion_n=1}
\end{align}
%%%%%%%%%%%%%%%%%%%%%%%%%%%%%%%%%%%%%%%%%%%%%%%%%%%
where $\delta_0(\bfx)$ is the linear density field. 

Provided a linear density field on grids as an initial condition, we calculate the nonlinear source terms given at the right-hand side of Eq.~(\ref{eq:recursion_formula}). The fast Fourier transform (FFT) facilitates the calculation of the derivative operators $\nabla_i$ which simply becomes a multiplication of $\bfk_i$ in Fourier space. We have presented details of the algorithm and implementation in Ref.~\cite{Taruya_Nishimichi_Jeong2018} (see their Sec.~II-C). Making use of the recursion relation in Eq.~(\ref{eq:recursion_formula}), we have previously generated the nonlinear density fields up to the fifth order, and studied both their morphological and statistical properties in a face-to-face comparison with $N$-body simulations. Other advantages of this method include that the evaluation of statistical quantities such as the power spectrum can be shared with the same grid-based measurement codes used to analyze the $N$-body simulations result, and that once the density fields are generated, the predictions can be scaled to any redshift analytically by using the scaling in Eq.~(\ref{eq:PT_expansion}).

Note cautiously that the single-stream PT treatment ceases to be adequate in the nonlinear regime where the multi-stream flow is generated, and it is recently shown that its impact on the prediction of matter distribution appear manifest even at large scales, and becomes significant as we go to higher order (e.g., \cite{Blas:2013aba,Bernardeau:2012ux,Nishimichi:2014rra,Nishimichi_etal2017}). In this respect, it is not trivial to answer whether the higher-order PT calculations improve the prediction of the covariance matrix or not (see also Ref.~\cite{Bertolini2016}). This is indeed one of our focuses in the present paper. We shall address this question by comparing the covariance matrix from \gridspt\, with that from a suite of $N$-body simulations.

%%%%%%%%%%%%%%%%%%%%%%%%%%%%%%%%%%%%%%%%%%%%%%%%%%%%%%%%%%%%%%%%%%%%%%%
%%%%%%%%%%%%%%%%%%%%%%%%%%%%%%%%%%%%%%%%%%%%%%%%%%%%%%%%%%%%%%%%%%%%%%%
\section{\gridspt\, calculation of the Covariance matrix}
\label{sec:GridSPT_covariance}
%%%%%%%%%%%%%%%%%%%%%%%%%%%%%%%%%%%%%%%%%%%%%%%%%%%%%%%%%%%%%%%%%%%%%%%
%%%%%%%%%%%%%%%%%%%%%%%%%%%%%%%%%%%%%%%%%%%%%%%%%%%%%%%%%%%%%%%%%%%%%%%

In this section, we present a perturbative calculation of the covariance matrix of the matter power spectrum. By using the grid-based SPT method, we include both the non-Gaussian contributions coming from the one-loop trispectrum and the effect of survey window function, in particular, coming from masking out some area (due to, for example, foreground objects such as bright stars or the Galactic plane).

%%--%%--%%--%%--%%--%%--%%--%%--%%--%%--%%--%%--%%--%%--%%--%%--%%--%%
%%--%%--%%--%%--%%--%%--%%--%%--%%--%%--%%--%%--%%--%%--%%--%%--%%--%%
\subsection{Preliminaries}
\label{subsec:preliminaries}
%%--%%--%%--%%--%%--%%--%%--%%--%%--%%--%%--%%--%%--%%--%%--%%--%%--%%
%%--%%--%%--%%--%%--%%--%%--%%--%%--%%--%%--%%--%%--%%--%%--%%--%%--%%

To model the density fields calculated from \gridspt\,, in what follows, we consider the density field in a comoving cube of the side length $L$. To begin with, we ignore the survey mask, and assume that the density field $\delta$ is defined everywhere on the grids. To deal with the density field defined on grids with discrete Fourier modes, we follow Ref.~\cite{dePutter_etal2012} and write down the the density field in Fourier space as: 
%%%%%%%%%%%%%%%%%%%%%%%%%%%%%%%%%%%%%%%%%%%%%%%%%%%%%%%%%%%%%%%%%%%%%%%
\begin{align}
 \delta_{\bfk} \equiv \frac{1}{V}\,\int_V d^3\bfx \,e^{i\,\bfk\cdot\bfx}\,\delta(\bfx)\,;\quad \bfk=\frac{2\pi}{L}\,\bfn
\end{align} 
%%%%%%%%%%%%%%%%%%%%%%%%%%%%%%%%%%%%%%%%%%%%%%%%%%%%%%%%%%%%%%%%%%%%%%%
with $V=L^3$ and $\bfn$ being the three-dimensional vector having integer components\footnote{For actual implementation of \gridspt, the integral over three-dimensional space is replaced with the summation over grid space, i.e., $\int d^3\bfx \,f(\bfx)\longrightarrow (V/N_{\rm grid})\,\sum_n\,f(x_n) $ with $N_{\rm grid}$ being the number of grids. }. Note that we define $\delta_{\bfk}$ to be dimensionless.

With the discrete Fourier modes, the power spectrum $P(k)$ is defined by
%%%%%%%%%%%%%%%%%%%%%%%%%%%%%%%%%%%%%%%%%%%%%%%%%%%%%%%%%%%%%%%%%%%%%%%
\begin{align}
 \langle\delta_{\bfk} \delta_{\bfk'}\rangle &= \frac{P(k)}{V}\,\deltaK_{\bfk+\bfk'}\,
\label{eq:def_pk_average}
\end{align}
%%%%%%%%%%%%%%%%%%%%%%%%%%%%%%%%%%%%%%%%%%%%%%%%%%%%%%%%%%%%%%%%%%%%%%%
where the symbol $\deltaK_{\bfk+\bfk'}$ represents the Kronecker delta. The bracket $\langle\cdots\rangle$ stands for the ensemble average over the infinite number of random density fields. For a single realization of density field, the monopole power spectrum is estimated by:
%%%%%%%%%%%%%%%%%%%%%%%%%%%%%%%%%%%%%%%%%%%%%%%%%%%%%%%%%%%%%%%%%%%%%%%
\begin{align}
 \hat{P}(k_i) \equiv \frac{V}{N_i} \,\sum_{|\bfk| \in k_i} |\delta_{\bfk}|^2,
\label{eq:PS_estimator}
\end{align}
%%%%%%%%%%%%%%%%%%%%%%%%%%%%%%%%%%%%%%%%%%%%%%%%%%%%%%%%%%%%%%%%%%%%%%%
where the summation is for the wavevectors $\bfk$ falling in a wavenumber bin labeled by $i$, $N_i$ is the number of Fourier modes in 
the bin, i.e., $N_i \simeq 4\pi k_i^2\Delta k/(2\pi/L)^3$, with $\Delta k$ being the bin width. Eq.~(\ref{eq:PS_estimator}) gives an unbiased estimation of the power spectrum, i.e., $\langle\hat{P}(k_i)\rangle=P(k_i)$, as long as the bin width is sufficiently small. The estimation of power spectrum in Eq.~(\ref{eq:PS_estimator}) adds a finite number of Fourier modes, each of which includes statistical fluctuations. The covariance matrix for the power spectrum estimator is then defined as
%%%%%%%%%%%%%%%%%%%%%%%%%%%%%%%%%%%%%%%%%%%%%%%%%%%%%%%%%%%%%%%%%%%%%%%
\begin{align}
 \cov[P(k_i),P(k_j)] &\equiv \Bigl\langle \hat{P}(k_i) \hat{P}(k_j)\Bigr\rangle - 
\Bigl\langle \hat{P}(k_i)\Bigr\rangle \Bigl\langle \hat{P}(k_j)\Bigr\rangle.
\label{eq:def_covariance}
\end{align}
%%%%%%%%%%%%%%%%%%%%%%%%%%%%%%%%%%%%%%%%%%%%%%%%%%%%%%%%%%%%%%%%%%%%%%%
Substituting Eq.~(\ref{eq:PS_estimator}) into the definition, we obtain 
%%%%%%%%%%%%%%%%%%%%%%%%%%%%%%%%%%%%%%%%%%%%%%%%%%%%%%%%%%%%%%%%%%%%%%%
\begin{align}
 \cov[P(k_i),P(k_j)]= 2\frac{\{P(k_i)\}^2}{N_i}\,\deltaK_{ij} + \frac{\overline{T}_{ij}}{V}.
\label{eq:covariance}
\end{align}
%%%%%%%%%%%%%%%%%%%%%%%%%%%%%%%%%%%%%%%%%%%%%%%%%%%%%%%%%%%%%%%%%%%%%%%
Here, the first term at the right-hand side is the diagonal covariance originated from the disconnected part of the four-point correlation, and hence called `Gaussian covariance'. On the other hand, the second term encodes the non-Gaussian contribution to the covariance matrix that in general exhibits non-vanishing off-diagonal components. Specifically, the non-Gaussian part comes from the connected part of the four-point correlation:
%%%%%%%%%%%%%%%%%%%%%%%%%%%%%%%%%%%%%%%%%%%%%%%%%%%%%%%%%%%%%%%%%%%%%%%
\begin{align}
 \overline{T}_{ij}\equiv \frac{1}{N_i}\sum_{|\bfk|\in k_i} \frac{1}{N_j}\sum_{|\bfk'|\in k_j} T(\bfk,-\bfk,\bfk',-\bfk')
\label{eq:averaged_trispectrum}
\end{align}
%%%%%%%%%%%%%%%%%%%%%%%%%%%%%%%%%%%%%%%%%%%%%%%%%%%%%%%%%%%%%%%%%%%%%%%
with the quantity $T$ being the trispectrum:
%%%%%%%%%%%%%%%%%%%%%%%%%%%%%%%%%%%%%%%%%%%%%%%%%%%%%%%%%%%%%%%%%%%%%%%
\begin{align}
 \langle\delta_{\bfk} \delta_{\bfk'} \delta_{\bfk''}\delta_{\bfk'''}\rangle_c &= \frac{T(\bfk,\bfk',\bfk'',\bfk''')}{V^3}\,\deltaK_{\bfk+\bfk'+\bfk''+\bfk'''},
\label{eq:def_trispectrum}
\end{align}
%%%%%%%%%%%%%%%%%%%%%%%%%%%%%%%%%%%%%%%%%%%%%%%%%%%%%%%%%%%%%%%%%%%%%%%
where the bracket $\langle\cdots\rangle_c$ implies the ensemble average subtracting the disconnected (Wick-contracted) part of the correlators.

The non-Gaussian contribution to the above covariance matrix expression may be further divided into two parts. One is the covariance arising from the nonlinear mode-coupling between the modes inside the survey region, called sub-survey modes ($k>2\pi/L$). The other part is called the {\it super-sample covariance}, originated from the coupling between the sub-survey modes and super-survey modes whose wavelengths exceeds the survey region ($k<2\pi/L$) \cite{Takada_Hu2013,Li_Hu_Takada2014b} (see also Refs.~\cite{Rimes_Hamilton2005,Hamilton_etal2006,Sefusatti_Crocce_Scoccimarro2006,Takada_Jain2009,Takahashi2009} for early works). The latter contribution is known to give an impact on the total covariance, and techniques to compute it have been developed using perturbation theory and $N$-body simulations \cite{Li_Hu_Takada2014a,Baldauf_etal2016}. In this paper, we do not consider the super-sample covariance, and rather focus on the non-Gaussian covariance between sub-survey modes\footnote{Strictly speaking, we consider a part of super-sample modes when we apply masks to account for the geometry of the survey volume in Sec.~\ref{subsec:mask}. To be more precise, we examine the covariance calculations with the survey masks shown in Fig.~\ref{fig:config_masks}, in which the Fourier modes with wavelength larger than the trimmed ``survey'' region, especially for sphere 1 and 2, are automatically considered up to the size of the parent cubic box.}. 
That is, we shall compare the \gridspt\, calculation with the covariance matrix estimated from sub-modes measured from a suite of $N$-body simulations. The quantitative estimation of the super-sample covariance with \gridspt~ is left for our future work. 

Note that the expressions given at Eqs.~(\ref{eq:covariance})-(\ref{eq:def_trispectrum}) are valid for the un-masked density fields. Taking the survey masks into account, the non-trivial mode coupling induced by the survey window function changes the structures of covariance. Consequently, even the Gaussian covariance produces non-vanishing off-diagonal components, which must be also accounted for in order to properly describe the covariance of observed density fields. We shall come back to this point in Sec.~\ref{subsec:survey_window}.

%%%%%%%%%%%%%%%%%%%%%%%%%%%%%%%%%%%%%%%%%%%%%%%%%%%%%%%%%%%%%%%%%%%%%%%
\begin{figure*}[tb]
%\vspace*{-2.8cm}
\begin{center}
%\hspace*{0.5cm}
\includegraphics[width=13cm,angle=0]{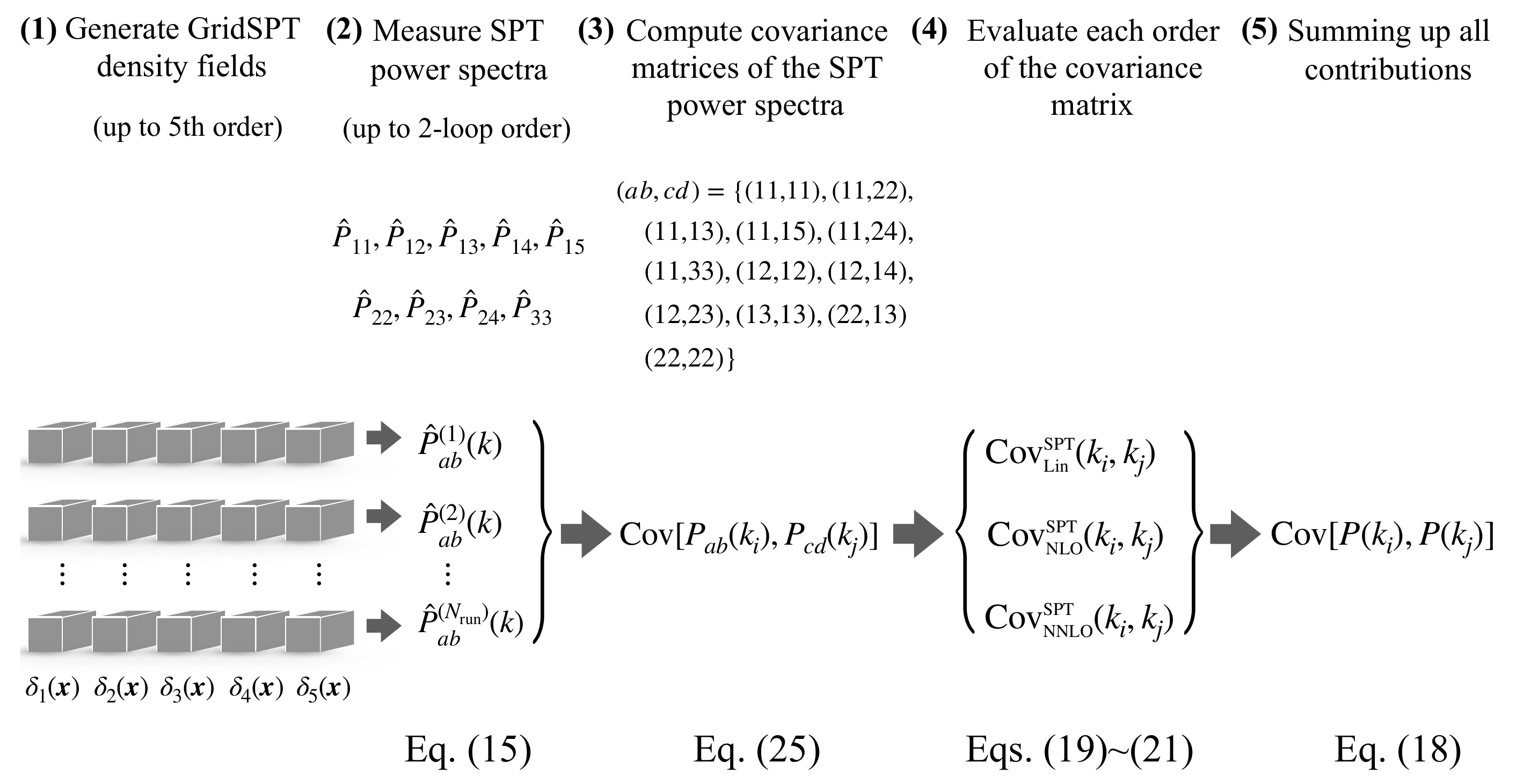}
\end{center}
\vspace*{-0.4cm}
\caption{Flow chart of the covariance estimation with \gridspt. For a perturbative calculation at NNLO, we first generate the PT density fields up to the fifth order. We repeat this to obtain $N_{\rm run}$ realizations, and for each realization, the SPT power spectra, $\hat{P}_{ab}$, are measured up to the two-loop order. These power spectra are used to estimate the covariance matrix, $\cov[P_{ab}(k_i),P_{cd}(k_j)]$, through  Eq.~(\ref{eq:estimator_covariance}), and plugging these covariance matrices into Eqs.~(\ref{eq:PT_covariance_Lin})-(\ref{eq:PT_covariance_NNLO}), the total covariance matrix is finally computed with Eq.~(\ref{eq:PT_expansion_covariance}).  This procedure is also applied to the cases with survey masks, just replacing the quantities with those evaluated with masked density fields, $\deltaw$ [see Eqs.~(\ref{eq:deltaw_1})-(\ref{eq:deltaw_5})]. 
\label{fig:flow_chart}
}
\end{figure*}
%%%%%%%%%%%%%%%%%%%%%%%%%%%%%%%%%%%%%%%%%%%%%%%%%%%%%%%%%%%%%%%%%%%%%%%

%%--%%--%%--%%--%%--%%--%%--%%--%%--%%--%%--%%--%%--%%--%%--%%--%%--%%
%%--%%--%%--%%--%%--%%--%%--%%--%%--%%--%%--%%--%%--%%--%%--%%--%%--%%
\subsection{The algorithm: Perturbative calculation of covariance matrix}
\label{subsec:non-Gaussian_covariance}
%%--%%--%%--%%--%%--%%--%%--%%--%%--%%--%%--%%--%%--%%--%%--%%--%%--%%
%%--%%--%%--%%--%%--%%--%%--%%--%%--%%--%%--%%--%%--%%--%%--%%--%%--%%

In this section, we shall present the algorithm for calculating covariance matrix from \gridspt\,. Perhaps the most obvious method in PT-based approaches (for example, in Ref.~\cite{Digvijay_Scoccimarro2019}) is to evaluate perturbatively the first and second terms of Eq.~(\ref{eq:covariance}) using the PT solutions. We could also use \gridspt\, to compute the required power spectrum and trispectrum. We shall, however, use an alternative method taking advantage of the fact that \gridspt\, generates multiple realizations of nonlinear density field. That is, we can directly estimate the ensemble average in Eq.~(\ref{eq:def_covariance}) by averaging over the \gridspt\, realizations. This method has a couple of advantages. First, we only need to measure the power spectrum, not the trispectrum, from each \gridspt\, realization. Second, the survey window function simply multiplies to the density field in configuration space, in contrast to the convolution required for the Fourier-space PT calculation.

Let us find the expression for the covariance matrix by substituting Eq.~(\ref{eq:PS_estimator}) into Eq.~(\ref{eq:def_covariance}):
%%%%%%%%%%%%%%%%%%%%%%%%%%%%%%%%%%%%%%%%%%%%%%%%%%%%%%%%%%%%%%%%%%%%%%%
\begin{widetext}
\begin{align}
& \cov[P(k_i),P(k_j)]= \frac{V}{N_i}\frac{V}{N_j}
%\nonumber
%\\
%&\times 
\sum_{|\bfk|\in k_i} \sum_{|\bfk'|\in k_j} \Bigl\{\bigl\langle\delta_{\bfk}\delta_{-\bfk}\delta_{\bfk'}\delta_{-\bfk'}\bigr\rangle - \bigl\langle\delta_{\bfk}\delta_{-\bfk}\bigr\rangle\,\bigl\langle\delta_{\bfk'}\delta_{-\bfk'}\bigr\rangle \Bigr\}. 
\end{align}
%%%%%%%%%%%%%%%%%%%%%%%%%%%%%%%%%%%%%%%%%%%%%%%%%%%%%%%%%%%%%%%%%%%%%%%
Applying the PT expansion given at Eq.~(\ref{eq:PT_expansion}), we obtain
%%%%%%%%%%%%%%%%%%%%%%%%%%%%%%%%%%%%%%%%%%%%%%%%%%%%%%%%%%%%%%%%%%%%%%%
\begin{align}
 \cov[P(k_i),P(k_j)]
&=\sum_{a,b,c,d}\,e^{(a+b+c+d)\eta}\,\frac{V}{N_i}\frac{V}{N_j}
\sum_{|\bfk|\in k_i} 
\sum_{|\bfk'|\in k_j} \Bigl\{\bigl\langle\delta_{a,\bfk}\delta_{b,-\bfk}\delta_{c,\bfk'}\delta_{d,-\bfk'}\bigr\rangle - \bigl\langle\delta_{a,\bfk}\delta_{b,-\bfk}\bigr\rangle\,\bigl\langle\delta_{c,\bfk'}\delta_{d,-\bfk'}\bigr\rangle \Bigr\} 
\nonumber
\\
&=\sum_{a,b,c,d}\,e^{(a+b+c+d)\eta}\,\frac{V}{N_i}\frac{V}{N_j}
\sum_{|\bfk|\in k_i} \sum_{|\bfk'|\in k_j} \Bigl\{\bigl\langle\mbox{Re}[\delta_{a,\bfk}\delta_{b,-\bfk}]\mbox{Re}[\delta_{c,\bfk'}\delta_{d,-\bfk'}]\bigr\rangle - \bigl\langle\mbox{Re}[\delta_{a,\bfk}\delta_{b,-\bfk}]\bigr\rangle\,\bigl\langle\mbox{Re}[\delta_{c,\bfk'}\delta_{d,-\bfk'}]\bigr\rangle \Bigr\}, 
\label{eq:cov_PT_expansion}
\end{align}
\end{widetext}
%%%%%%%%%%%%%%%%%%%%%%%%%%%%%%%%%%%%%%%%%%%%%%%%%%%%%%%%%%%%%%%%%%%%%%%
where in the second equality we used the fact that the summation $\sum_{|\bfk|\in k_i}$ takes all Fourier modes inside the spherical shell defined by the bin $k_i$ ($| |\bfk| -k_i|<\Delta k/2$) and the reality condition of $\delta({\bm r})$: $\delta(-\bfk) = \delta^*(\bfk)$.
We can further simplify Eq.~(\ref{eq:cov_PT_expansion}) by defining the following estimator for the cross power spectrum of the $a$-th order density field and the $b$-th order density field
%%%%%%%%%%%%%%%%%%%%%%%%%%%%%%%%%%%%%%%%%%%%%%%%%%%%%%%%%%%%%%%%%%%%%%%
\begin{align}
\hat{P}_{ab}(k_i)\equiv \frac{V}{N_i}\sum_{|\bfk|\in k_i}\,\mbox{Re}[\delta_{a,\bfk}\delta_{b,-\bfk}],
\label{eq:estimator_Pab}
\end{align}
%%%%%%%%%%%%%%%%%%%%%%%%%%%%%%%%%%%%%%%%%%%%%%%%%%%%%%%%%%%%%%%%%%%%%%%
which is an unbiased estimator for the SPT power spectrum $P_{ab}(k)$, $\langle\hat{P}_{ab}(k)\rangle=P_{ab}(k)$, where the power spectrum $P_{ab}$ is defined by
%%%%%%%%%%%%%%%%%%%%%%%%%%%%%%%%%%%%%%%%%%%%%%%%%%%%%%%%%%%%%%%%%%%%%%%
\begin{align}
%\bigl\langle\mbox{Re}[\delta_{a,\bfk}\delta_{b,\bfk'}]\bigr\rangle = \frac{P_{ab}(k)}{V}\,\deltaK_{\bfk+\bfk'}. 
\frac{1}{2}\bigl\langle \delta_{a,\bfk}\delta_{b,\bfk'}+\delta_{a,\bfk'}\delta_{b,\bfk}\bigr\rangle = \frac{P_{ab}(k)}{V}\,\deltaK_{\bfk+\bfk'}. 
\label{eq:pk_SPT}
\end{align}
%%%%%%%%%%%%%%%%%%%%%%%%%%%%%%%%%%%%%%%%%%%%%%%%%%%%%%%%%%%%%%%%%%%%%%%
Using Eq.(\ref{eq:pk_SPT}), Eq.~(\ref{eq:cov_PT_expansion}) is rewritten in a simple form: 
%%%%%%%%%%%%%%%%%%%%%%%%%%%%%%%%%%%%%%%%%%%%%%%%%%%%%%%%%%%%%%%%%%%%%%%
\begin{widetext}
\begin{align} 
 \cov[P(k_i),P(k_j)]
&=\sum_{a,b,c,d}\,e^{(a+b+c+d)\eta}\, 
\Bigl\{\bigl\langle \hat{P}_{ab}(k_i) \hat{P}_{cd}(k_i)\bigr\rangle - \bigl\langle \hat{P}_{ab}(k_i)\bigr\rangle \bigl\langle\hat{P}_{cd}(k_i)\bigr\rangle\Bigr\}
\nonumber
\\
&\equiv \sum_{a,b,c,d}\,e^{(a+b+c+d)\eta}\,\cov[P_{ab}(k_i),\,P_{cd}(k_j)].
\label{eq:covpk_PT}
\end{align}
\end{widetext}
%%%%%%%%%%%%%%%%%%%%%%%%%%%%%%%%%%%%%%%%%%%%%%%%%%%%%%%%%%%%%%%%%%%%%%%

Eq.~(\ref{eq:covpk_PT}) is a general perturbative expression for the covariance matrix. Recalling further that the $n$-th order PT density field, $\delta_{n,\bfk}$, is the quantity of the order of $\mathcal{O}(\delta_1^n)$, the expansion form given above is reorganized under the Gaussian initial condition as follows:
%%%%%%%%%%%%%%%%%%%%%%%%%%%%%%%%%%%%%%%%%%%%%%%%%%%%%%%%%%%%%%%%%%%%%%%
\begin{widetext}
\begin{align}
 \cov[P(k_i),P(k_j)] & = e^{4\,\eta} \,\covSPTL(k_i,k_j)
+ e^{6\,\eta} \,\covSPTNLO(k_i,k_j)
+ e^{8\,\eta} \,\covSPTNNLO(k_i,k_j)+\cdots.
\label{eq:PT_expansion_covariance}
\end{align}
%%%%%%%%%%%%%%%%%%%%%%%%%%%%%%%%%%%%%%%%%%%%%%%%%%%%%%%%%%%%%%%%%%%%%%%
Here, the first term at right-hand-side, $\covSPTL$, represents the linear-order covariance. 
The two other terms, i.e., $\covSPTNLO$, and $\covSPTNNLO$, represent the higher-order contributions, which we respectively denote by the next-to-leading order (NLO) and next-to-next-to-leading order (NNLO) covariance matrices. Their explicit expressions are given as follows:
%%%%%%%%%%%%%%%%%%%%%%%%%%%%%%%%%%%%%%%%%%%%%%%%%%%%%%%%%%%%%%%%%%%%%%%
\begin{align}
 \covSPTL(k_i,k_j) &= \cov[P_{11}(k_i),\,P_{11}(k_j)],
\label{eq:PT_covariance_Lin}
%\end{align}
%%%%%%%%%%%%%%%%%%%%%%%%%%%%%%%%%%%%%%%%%%%%%%%%%%%%%%%%%%%%%%%%%%%%%%%
\\
%%%%%%%%%%%%%%%%%%%%%%%%%%%%%%%%%%%%%%%%%%%%%%%%%%%%%%%%%%%%%%%%%%%%%%%
%\begin{align}
 \covSPTNLO(k_i,k_j) &= 
 \Bigl\{ 
    \cov[P_{11}(k_i),P_{22}(k_j)] 
+2\,\cov[P_{11}(k_i),P_{13}(k_j)] 
%\nonumber
%\\
%&\quad 
+(i \leftrightarrow j)\Bigr\} 
+4\,\cov[P_{12}(k_i),P_{12}(k_j)],
\label{eq:PT_covariance_NLO}
%\end{align}
%%%%%%%%%%%%%%%%%%%%%%%%%%%%%%%%%%%%%%%%%%%%%%%%%%%%%%%%%%%%%%%%%%%%%%%
\\
%%%%%%%%%%%%%%%%%%%%%%%%%%%%%%%%%%%%%%%%%%%%%%%%%%%%%%%%%%%%%%%%%%%%%%%
%\begin{align}
\covSPTNNLO(k_i,k_j)& =  
\,\Bigl\{
  2\,\cov[P_{11}(k_i),P_{15}(k_j)]
+ 2\,\cov[P_{11}(k_i),P_{24}(k_j)]
%\nonumber
%\\
%&\quad
+4\,\cov[P_{12}(k_i),P_{14}(k_j)] 
\nonumber
\\
&\quad
 +   \cov[P_{11}(k_i),P_{33}(k_j)] 
 +2\,\cov[P_{22}(k_i),P_{13}(k_j)] 
 +4\,\cov[P_{12}(k_i),P_{23}(k_j)]
%\nonumber
%\\
%&\quad
 + (i \leftrightarrow j)\,\Bigr\}
\nonumber
\\
&\quad
 +4\,\cov[P_{13}(k_i),P_{13}(k_j)] 
 +\cov[P_{22}(k_i),P_{22}(k_j)]. 
\label{eq:PT_covariance_NNLO}
\end{align}
\end{widetext}
%%%%%%%%%%%%%%%%%%%%%%%%%%%%%%%%%%%%%%%%%%%%%%%%%%%%%%%%%%%%%%%%%%%%%%%
Here, we denote by $(i \leftrightarrow j)$ the terms obtained by exchanging the two indices, $i$ and $j$, in those preceding in the brace. Note that in deriving the expressions, we have used the symmetry of $P_{ab}=P_{ba}$.

Eq.~(\ref{eq:PT_expansion_covariance}) with Eqs.~(\ref{eq:PT_covariance_Lin})-(\ref{eq:PT_covariance_NNLO}) provides the basis for calculating the covariance matrix with \gridspt. To clarify their statistical meanings, we rewrite each contribution of Eqs.~(\ref{eq:PT_covariance_Lin})-(\ref{eq:PT_covariance_NNLO}) in terms of the power spectrum and trispectrum, similarly to Eq.~(\ref{eq:covariance}):
%%%%%%%%%%%%%%%%%%%%%%%%%%%%%%%%%%%%%%%%%%%%%%%%%%%%%%%%%%%%%%%%%%%%%%%
\begin{align}
 \covSPTL(k_i,k_j)& =  2\frac{\{P_{11}(k_i)\}^2}{N_i}\,\deltaK_{ij},
\label{eq:covSPT_Lin_2nd}
\\
 \covSPTNLO(k_i,k_j)& =  \frac{4}{N_i}\,\deltaK_{ij} \, P_{11}(k_i)\bigl\{P_{22}(k_i)+2\,P_{13}(k_i)\bigr\} 
\nonumber\\
&+ \frac{\overline{T}_{ij}^{\rm tree}}{V},
\label{eq:covSPT_NLO_2nd}
\\
 \covSPTNNLO(k_i,k_j)& =  \frac{4}{N_i}\,\deltaK_{ij}\,\Bigl[P_{11}(k_i)\bigl\{2\,P_{15}(k_i)+2\,P_{24}(k_i)
\nonumber
\\
&+P_{33}(k_i)\bigr\}+2P_{13}(k_i)\bigl\{P_{13}(k_i)+P_{22}(k_i)\bigr\}
\nonumber
\\
&+\frac{1}{2}\{P_{22}(k_i)\}\Bigr]\deltaK_{ij}+\frac{\overline{T}_{ij}^{\rm 1\mbox{-}loop}}{V}, 
\label{eq:covSPT_NNLO_2nd}
\end{align}
%%%%%%%%%%%%%%%%%%%%%%%%%%%%%%%%%%%%%%%%%%%%%%%%%%%%%%%%%%%%%%%%%%%%%%%
where the matrices $\overline{T}_{ij}^{\rm tree}$ and $\overline{T}_{ij}^{\rm1\mbox{-}loop}$ are respectively the non-Gaussian contributions arising from the tree-level and one-loop (NLO) trispectrum, given in Eq.~(\ref{eq:averaged_trispectrum}). Thus, the off-diagonal part of $\covSPTNLO$ and $\covSPTNNLO$ represents the non-Gaussian covariance coming from the connected trispectrum, while the diagonal components are the mixture of Gaussian and non-Gaussian contributions. We emphasize again that the \gridspt\, implementation allows us to calculate the off-diagonal component of the covariant matrix without explicitly computing the one-loop trispectrum.

Now, the procedure to compute the covariance with \gridspt\, up to the NNLO (i.e., including the trispectrum at one-loop order) is summarized as follows. First, we generate a large number ($N_{\rm run}$) of nonlinear density field with \gridspt, and measure all possible SPT power spectra $\hat{P}_{ab}$ up to the two-loop order ($\hat{P}_{ab}\propto\mathcal{O}(\delta_1^6)$) for each realization. Repeating the power spectrum measurements over all realizations, we next evaluate the covariance matrices in Eqs.~(\ref{eq:PT_covariance_Lin})-(\ref{eq:PT_covariance_NNLO}), for which we adopt the following estimator: 
%%%%%%%%%%%%%%%%%%%%%%%%%%%%%%%%%%%%%%%%%%%%%%%%%%%%%%%%%%%%%%%%%%%%%%%
\begin{align}
& \cov[P_{ab}(k_i),P_{cd}(k_j)] = \frac{1}{N_{\rm run}-1} 
\nonumber
\\
&\quad \times
\sum_{n=1}^{N_{\rm run}} \Bigl\{\hat{P}_{ab}^{(n)}(k_i)-
\overline{P}_{ab}
(k_i)\Bigr\}
\Bigl\{\hat{P}_{cd}^{(n)}(k_j)-
\overline{P}_{cd}
(k_j)\Bigr\}.
\label{eq:estimator_covariance}
\end{align}
%%%%%%%%%%%%%%%%%%%%%%%%%%%%%%%%%%%%%%%%%%%%%%%%%%%%%%%%%%%%%%%%%%%%%%%
Here, $N_{\rm run}$ is the number of realizations, and the estimator $\hat{P}_{ab}^{(n)}$ represents the SPT power spectra measured from the $n$-th realization. The quantity $\overline{P}_{ab}$ is the SPT spectrum averaged all realizations, given by  
%%%%%%%%%%%%%%%%%%%%%%%%%%%%%%%%%%%%%%%%%%%%%%%%%%%%%%%%%%%%%%%%%%%%%%%
\begin{align}
& \overline{P}_{ab}(k_i)=\frac{1}{N_{\rm run}} \sum_{n=1}^{N_{\rm run}}\hat{P}_{ab}^{(n)}(k_i).
\end{align}
%%%%%%%%%%%%%%%%%%%%%%%%%%%%%%%%%%%%%%%%%%%%%%%%%%%%%%%%%%%%%%%%%%%%%%%
Summing up all the contributions, the leading and higher-order covariance matrices, $\covSPTL$, $\covSPTNLO$, $\covSPTNNLO$ are computed, and the total covariance is finally obtained from Eq.~(\ref{eq:PT_expansion_covariance}). 

We summarize the procedure as the flow chart in Fig.~\ref{fig:flow_chart}.

%%--%%--%%--%%--%%--%%--%%--%%--%%--%%--%%--%%--%%--%%--%%--%%--%%--%%
%%--%%--%%--%%--%%--%%--%%--%%--%%--%%--%%--%%--%%--%%--%%--%%--%%--%%
\subsection{Survey window function and mask}
\label{subsec:survey_window}
%%--%%--%%--%%--%%--%%--%%--%%--%%--%%--%%--%%--%%--%%--%%--%%--%%--%%
%%--%%--%%--%%--%%--%%--%%--%%--%%--%%--%%--%%--%%--%%--%%--%%--%%--%%

So far, we have considered the covariance matrix without the survey window function effects. However, with the configuration-space density field from \gridspt, it is rather straightforward to incorporate the survey window function effect into the PT density fields as a post process. Also, the covariance matrix calculation outlined in Sec.~\ref{subsec:non-Gaussian_covariance} is general enough to be applicable to the window-function convolved density field without any modification.

One subtlety arising from a survey window function is that the window function breaks the homogeneity of the survey volume. Therefore, the volume average of the density field convolved with the window function generally differs from the true ensemble mean (e.g., Ref.~\cite{dePutter_etal2012}). This means that we must exploit the density estimator in order to preserve the properties of the underlying density field. As for the definite example of the window function, in this paper, we shall consider cases where some part of the survey volume is masked out. But, the analysis method below holds for general window functions.

Denoting the window function characterizing the survey masks by $W(\bfx)$, we consider the following density estimator, $\deltaw$:
%%%%%%%%%%%%%%%%%%%%%%%%%%%%%%%%%%%%%%%%%%%%%%%%%%%%%%%%%%%%%%%%%%%%%%%
\begin{align}
 \deltaw(\bfx)\equiv \frac{W(\bfx)\,\rho(\bfx)}{(1/V_{\rm w})\,\int_V\,d^3\bfx\,W(\bfx)\,\rho(\bfx)} - W(\bfx),
\label{eq:def_deltaw}
\end{align}
%%%%%%%%%%%%%%%%%%%%%%%%%%%%%%%%%%%%%%%%%%%%%%%%%%%%%%%%%%%%%%%%%%%%%%%
where $\rho$ is the true mass or number density field given by $\rho(\bfx)=\overline{\rho}\,\{1+\delta(\bfx)\}$, with the density fluctuation $\delta$ having zero mean. The volume $V_{\rm w}$ represents the actual survey volume defined by
%%%%%%%%%%%%%%%%%%%%%%%%%%%%%%%%%%%%%%%%%%%%%%%%%%%%%%%%%%%%%%%%%%%%%%%
\begin{align}
 V_{\rm w}\equiv \int_V d^3\bfx\,W(\bfx),
 \label{eq:def_VW}
\end{align}
%%%%%%%%%%%%%%%%%%%%%%%%%%%%%%%%%%%%%%%%%%%%%%%%%%%%%%%%%%%%%%%%%%%%%%%
which differs from the entire cubic volume $V$. Note that taking the volume average, the density fluctuation defined above leads to 
%%%%%%%%%%%%%%%%%%%%%%%%%%%%%%%%%%%%%%%%%%%%%%%%%%%%%%%%%%%%%%%%%%%%%%%
\begin{align}
\int_V d^3\bfx \,\deltaw(\bfx) =0.
\end{align}
%%%%%%%%%%%%%%%%%%%%%%%%%%%%%%%%%%%%%%%%%%%%%%%%%%%%%%%%%%%%%%%%%%%%%%%

Given the density estimator above, a perturbative calculation of the covariance matrix, as described in Sec.~\ref{subsec:non-Gaussian_covariance}, is made with the PT expansion of $\deltaw$. Using the true density fluctuation $\delta$, we rewrite Eq.~(\ref{eq:def_deltaw}) as
%%%%%%%%%%%%%%%%%%%%%%%%%%%%%%%%%%%%%%%%%%%%%%%%%%%%%%%%%%%%%%%%%%%%%%%
\begin{align}
\deltaw(\bfx)=\frac{W(\bfx)
\left(\delta(\bfx)-\dellocal\right)}{1+\dellocal}
=
W(\bfx)
\tilde{\delta}(\bfx)
\sum_{n=0}^{\infty}
(-\dellocal)^n
,
\label{eq:estimator_delta_with_mask}
\end{align}
%%%%%%%%%%%%%%%%%%%%%%%%%%%%%%%%%%%%%%%%%%%%%%%%%%%%%%%%%%%%%%%%%%%%%%%
where the quantity $\dellocal$ is the local mean of the density fluctuation, given by
%%%%%%%%%%%%%%%%%%%%%%%%%%%%%%%%%%%%%%%%%%%%%%%%%%%%%%%%%%%%%%%%%%%%%%%
\begin{align}
 \dellocal\equiv \frac{1}{V_{\rm w}}\,\int_{V} d^3\bfx\,W(\bfx)\,\delta(\bfx),
\end{align}
%%%%%%%%%%%%%%%%%%%%%%%%%%%%%%%%%%%%%%%%%%%%%%%%%%%%%%%%%%%%%%%%%%%%%%%
and $\tilde{\delta}(\bfx)\equiv \delta(\bfx)-\dellocal$.
Expanding Eq.~(\ref{eq:estimator_delta_with_mask}), we compute perturbatively the density field at each order. Note that the local mean $\overline{\delta}$ is a statistically fluctuating quantity that varies realization by realization, and we have to expand both the true density and local density fields, $\delta$ and $\overline{\delta}$. Writing the expansion form of $\deltaw$ as $\deltaw(\bfx) =\sum_n e^{n\,\eta}\,\delta_{{\rm w},n}(\bfx)$, we obtain the expressions of $\delta_{{\rm w},n}$ up to the fifth order:
%%%%%%%%%%%%%%%%%%%%%%%%%%%%%%%%%%%%%%%%%%%%%%%%%%%%%%%%%%%%%%%%%%%%%%%
\begin{align}
 \delta_{{\rm w},1}(\bfx)= & W(\bfx)\tilde{\delta}_1,
\label{eq:deltaw_1}
\\
 \delta_{{\rm w},2}(\bfx)= & W(\bfx)\bigl\{\tilde{\delta}_2 -\dellocal_1 \tilde{\delta}_1\bigr\},
\label{eq:deltaw_2}
\\
 \delta_{{\rm w},3}(\bfx)= & W(\bfx)\bigl\{ \tilde{\delta}_3 -\dellocal_1 \tilde{\delta}_2+ (-\dellocal_2+\dellocal_1^2)\tilde{\delta}_1\bigr\},
\label{eq:deltaw_3}
\\
 \delta_{{\rm w},4}(\bfx)= & W(\bfx)\bigl\{ \tilde{\delta}_4  -\dellocal_1 \tilde{\delta}_3 + (-\dellocal_2+\dellocal_1^2)\tilde{\delta}_2 
\nonumber
\\
&+ (-\dellocal_3+2\dellocal_1\dellocal_2-\dellocal_1^3)\tilde{\delta}_1 
\bigr\}
\label{eq:deltaw_4}
\\
 \delta_{{\rm w},5}(\bfx)= & W(\bfx)\bigl\{ \tilde{\delta}_5  -\dellocal_1 \tilde{\delta}_4 + (-\dellocal_2+\dellocal_1^2)\tilde{\delta}_3 
\nonumber
\\
& + (-\dellocal_3+2\dellocal_1\dellocal_2-\dellocal_1^3)\tilde{\delta}_2
\nonumber
\\
&  +(-\dellocal_4+2\dellocal_1\dellocal_3+\dellocal_2^2-3\dellocal_1^2\dellocal_2+\dellocal_1^4)\tilde{\delta}_1
\bigr\},
\label{eq:deltaw_5}
\end{align}
%%%%%%%%%%%%%%%%%%%%%%%%%%%%%%%%%%%%%%%%%%%%%%%%%%%%%%%%%%%%%%%%%%%%%%%
where the subscript indicates the perturbation-theory order of the quantity.

Note that as it is the density contrast averaged over the survey volume, the numerical value of $\dellocal$ is typically very small. However, that does not guarantee that the actual impact of the $\dellocal$ in Eqs.~(\ref{eq:deltaw_2})-(\ref{eq:deltaw_5}) on the covariance matrix is negligible \cite{dePutter_etal2012}. For example, Ref.~\cite{Digvijay_Scoccimarro2019} have shown that while its contribution to the power spectrum is small, the local average (i.e., $\dellocal$) contributes non-negligibly to the covariance matrix.

%%%%%%%%%%%%%%%%%%%%%%%%%%%%%%%%%%%%%%%%%%%%%%%%%%%%%%%%%%%%%%%%%%%%%%%
\begin{figure*}[tb]
%\vspace*{-2.8cm}
\begin{center}
%\hspace*{0.5cm}
\includegraphics[width=15cm,angle=0]{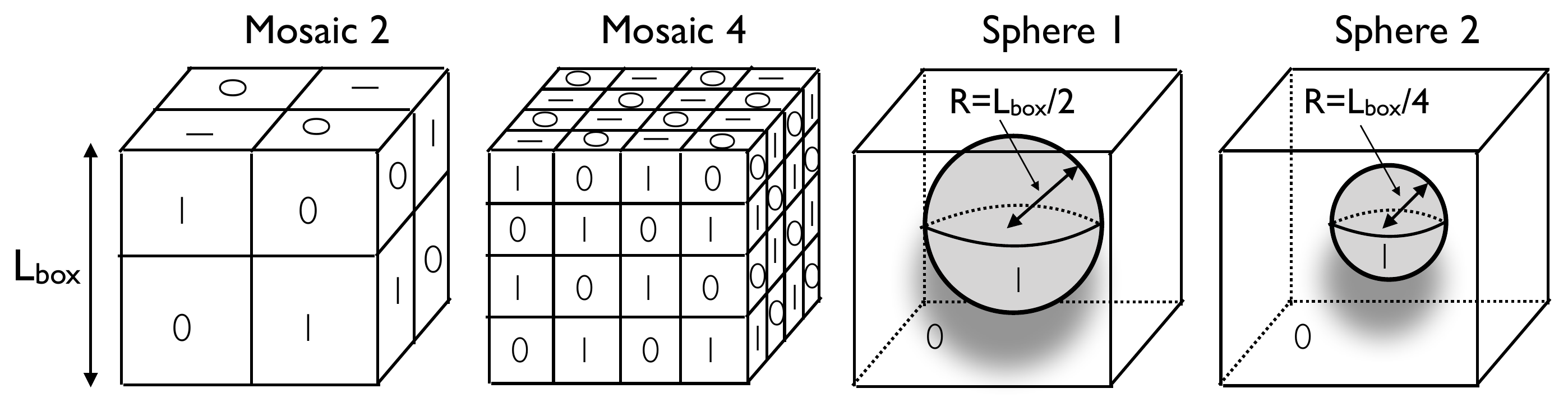}
\end{center}
\vspace*{-0.4cm}
\caption{Setup of survey masks considered in this paper. Here, the labels '1' and '0' indicate the window function of $W(\bfx)=1$ and $0$, respectively. 
\label{fig:config_masks}
}
\end{figure*}
%%%%%%%%%%%%%%%%%%%%%%%%%%%%%%%%%%%%%%%%%%%%%%%%%%%%%%%%%%%%%%%%%%%%%%%

Because \gridspt\, output is $\delta_n(\bfx)$ at each order, we can easily evaluate the right hand sides of Eqs.~(\ref{eq:deltaw_1})-(\ref{eq:deltaw_5}) to obtain the density field $\delta_{w,n}$ for a given survey window function. Then, the implementation for the covariance matrix goes parallel to the case without the window function, following the flow chart in Fig.~\ref{fig:flow_chart}. That is, from the \gridspt\, output, we first obtain $\delta_{{\rm w},n}(\bfx)$ in real space, then measure the SPT power spectra for the masked density fields, $\hat{P}_{{\rm w},ab}$, through Eq.~(\ref{eq:estimator_Pab}). Repeating the measurements of $P_{{\rm w},ab}$ over the $N_{\rm run}$ realizations, the covariance matrix of $P_{{\rm w},ab}$ is computed with Eq.~(\ref{eq:estimator_covariance}), and the covariance up to the NNLO is evaluated according to Eq.~(\ref{eq:PT_expansion_covariance}) with Eq.~(\ref{eq:PT_covariance_Lin})-(\ref{eq:PT_covariance_NNLO}).

In contrast to the case without window function, however, the expressions given at Eqs.~(\ref{eq:covSPT_Lin_2nd})-(\ref{eq:covSPT_NNLO_2nd}) are no longer adequate due to the non-trivial mode coupling arising from the window function, which leads to the non-vanishing off-diagonal components in the disconnected or Gaussian covariance. Therefore, in order to calculate the covariance matrix from the four-point correlators, one also needs to include $\covSPTL$ to correctly account for the off-diagonal components of the covariance matrix, in addition to the higher-order contributions given at Eqs.~(\ref{eq:PT_covariance_NLO}) and (\ref{eq:PT_covariance_NNLO}). The \gridspt\, implementation bypasses this complexity as we can estimate the covariance matrix of the power spectra from the multiple random realizations.

Finally, for the real galaxy surveys where the expected mean number density varies over the survey volume due to, for example, survey selection function, the estimator given at Eq.~(\ref{eq:def_deltaw}) or (\ref{eq:estimator_delta_with_mask}) is not optimal. Rather, the use of the FKP estimator \cite{FKP} would be an optimal choice, and when the Gaussian covariance dominates, it is shown to give a minimum-variance estimator.  Indeed, Ref.~\cite{Digvijay_Scoccimarro2019} adopted this estimator to analytically compute the power-spectrum covariance. Since the main purpose of this paper is to demonstrate explicitly the covariance calculation with \gridspt\, and to compare the higher-order predictions with $N$-body simulation, we shall below stick to a simple estimator at Eq.~(\ref{eq:def_deltaw}). The \gridspt\, calculation of covariance matrix with FKP or other optimal estimators is straightforward.

%%%%%%%%%%%%%%%%%%%%%%%%%%%  TABLE  %%%%%%%%%%%%%%%%%%%%%%%%%%%%%%%
\begin{table}[tb]
\caption{\label{tab:nbody_gridSPT} Parameter setup for $N$-body 
simulations and {\tt GridSPT}. The name {\tt GridSPT}-1 and {\tt GridSPT}-2 respectively imply the grid-based SPT simulations without and with survey masks.  }
\begin{ruledtabular}
\begin{tabular}{lccc}
Name & $L_{\rm box}$ & \# of particles/grids & \# of runs  
\\
\hline
$N$-body & $512\,h^{-1}$\,Mpc & $256^3$ particles & $10,000$  
\\
\hline
{\tt GridSPT}-1  & $512\,h^{-1}$Mpc & $256^3$ grids & $100,000$
\\
\hline
{\tt GridSPT}-2 & $512\,h^{-1}$Mpc & $256^3$ grids & $50,000$
\end{tabular}
\end{ruledtabular}
\end{table}
%%%%%%%%%%%%%%%%%%%%%%%%%%%  TABLE  %%%%%%%%%%%%%%%%%%%%%%%%%%%%%%%

%%%%%%%%%%%%%%%%%%%%%%%%%%%%%%%%%%%%%%%%%%%%%%%%%%%%%%%%%%%%%%%%%%%%%%%
%%%%%%%%%%%%%%%%%%%%%%%%%%%%%%%%%%%%%%%%%%%%%%%%%%%%%%%%%%%%%%%%%%%%%%%
\section{Results}
\label{sec:results}
%%%%%%%%%%%%%%%%%%%%%%%%%%%%%%%%%%%%%%%%%%%%%%%%%%%%%%%%%%%%%%%%%%%%%%%
%%%%%%%%%%%%%%%%%%%%%%%%%%%%%%%%%%%%%%%%%%%%%%%%%%%%%%%%%%%%%%%%%%%%%%%

We are in a position to present the results of the covariance estimation with \gridspt. In this section, focusing mainly on the non-Gaussian contributions, we shall present a detailed comparison between the covariance matrices obtained from the \gridspt\, and those measured from the $N$-body simulations. After describing the setup of simulations and \gridspt\, calculations in Sec.~\ref{subsec:setup}, we shall present the results with and without survey window function, respectively, in Sec.~\ref{subsec:no_mask} and \ref{subsec:mask}.

%%--%%--%%--%%--%%--%%--%%--%%--%%--%%--%%--%%--%%--%%--%%--%%--%%--%%
%%--%%--%%--%%--%%--%%--%%--%%--%%--%%--%%--%%--%%--%%--%%--%%--%%--%%
\subsection{Setup}
\label{subsec:setup}
%%--%%--%%--%%--%%--%%--%%--%%--%%--%%--%%--%%--%%--%%--%%--%%--%%--%%
%%--%%--%%--%%--%%--%%--%%--%%--%%--%%--%%--%%--%%--%%--%%--%%--%%--%%

As for the fiducial cosmological model, we use the flat-$\Lambda$CDM model with the cosmological parameters determined by Planck 2015 \cite{Planck2015_XIII}: $\Omega_{\rm m}=0.3156$ for the matter density, $\Omega_\Lambda=0.6844$ for the dark energy density with equation-of-state parameter $w=-1$, $\Omega_{\rm b}/\Omega_{\rm m}=0.1558$ for the baryon fraction, $n_s=0.9645$ for the scalar spectral index, $h=0.6727$ for the Hubble parameter, and finally, $A_s=2.2065\times10^{-9}$ for the amplitude of primordial scalar power spectrum: $P_s(k)=A_s(k/0.05\,$Mpc$)^{n_s}$.

%%%%%%%%%%%%%%%%%%%%%%%%%%%%%%%%%%%%%%%%%%%%%%%%%%%%%%%%%%%%%%%%%%%%%%%
\begin{figure*}[tb]
\vspace*{-0.8cm}
\begin{center}
%\hspace*{0.5cm}
\includegraphics[width=13cm,angle=0]{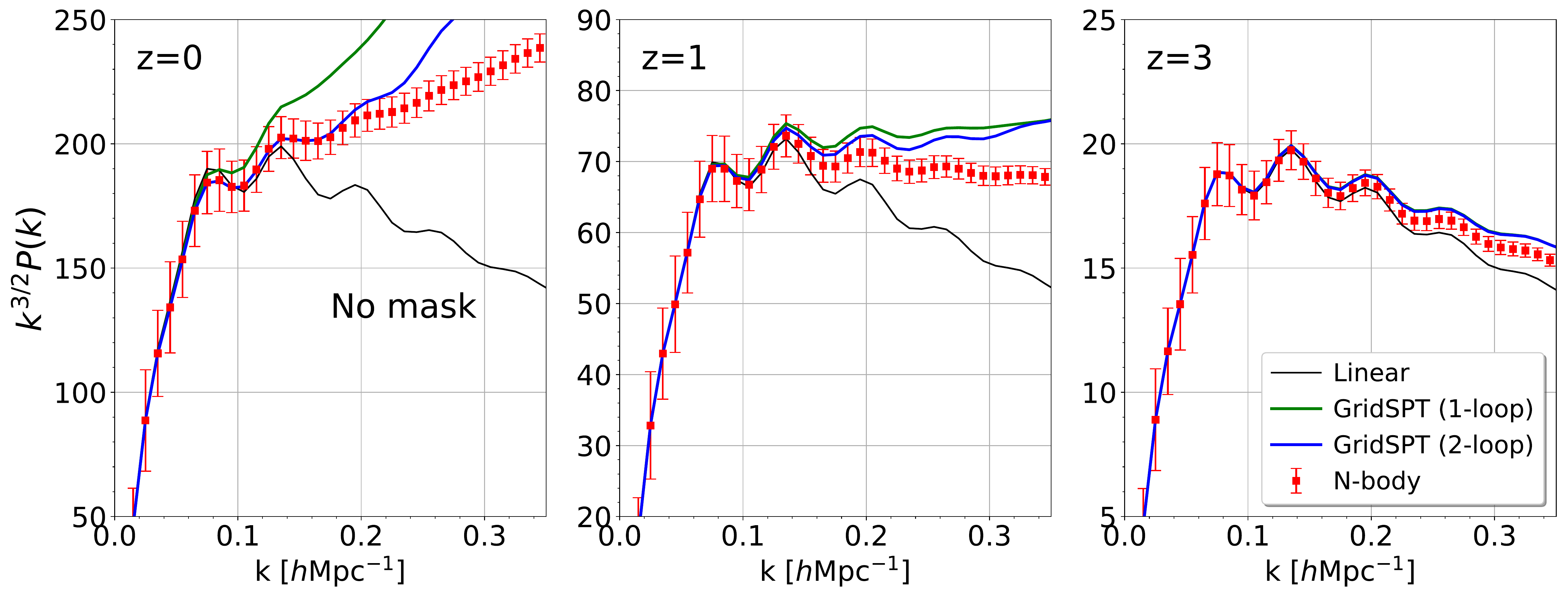}
\end{center}
\vspace*{-0.5cm}
\caption{Power spectra obtained from the {\tt GridSPT} calculations and $N$-body simulations at $z=0$ (left), $1$ (middle), and $3$ (right). For \gridspt, the results averaged over $100,000$ realizations are shown. While the black solid lines are the linear-order power spectra, the green and blue solid lines represent the results at one-loop (NLO) and two-loop order (NNLO). The $N$-body results are shown in filled red squares, which are obtained by averaging over $10,000$ simulations. The associated errorbars indicate the standard deviation. 
\label{fig:pk_gridSPT_no_mask}
}
\end{figure*}
%%%%%%%%%%%%%%%%%%%%%%%%%%%%%%%%%%%%%%%%%%%%%%%%%%%%%%%%%%%%%%%%%%%%%%%

The setup of $N$-body simulations and {\tt GridSPT} calculations are summarized in Table \ref{tab:nbody_gridSPT}. The cosmological $N$-body simulations are carried out by the publicly available code, {\tt GADGET-2} \cite{Springel:2005mi}, with the initial density field calculated with a code developed in Ref.~\cite{Nishimichi:2008ry} and parallelized in Ref.~\cite{Valageas:2010yw} based on the second-order Lagrangian perturbation theory (2LPT; \cite{Scoccimarro1998,Crocce:2006ve}). To make a robust estimation of the power spectrum covariance, a large number of realizations are necessary. Since our main focus is to test and validate the covariance estimation with \gridspt\, on large scales, we decided to run low-resolution simulations (i.e., the cubic box of the side length $L_{\rm box}=512\,h^{-1}$\,Mpc with $256^3$ particles, the Nyquist frequency of $k_{\rm Ny}=1.57\,h/{\rm Mpc}$) to reduce the cost and disk space, and we have performed $10,000$ independent random realizations with the output redshifts $z=0$, $1$, and $3$. This resolution is enough to study the power spectra for $k<0.3\,h/{\rm Mpc}$ \cite{Jeong:2006xd,Nishimichi_etal2017}. For \gridspt, taking advantage of the FFT, a much faster calculation is possible with the same resolution as in the $N$-body simulations. We have, in the end, created $50,000$ and $100,000$ realizations in the cases, respectively, with and without the survey window function\footnote{For reference, with the CPU of Xeon E5-2695 2.1GHz and using the 36 threads of FFT, it takes roughly 20 seconds to generate a single realization data (this includes the power spectrum calculations). Taking the survey masks at each order into account, it takes 30 seconds. }. For all analyses, we bin the Fourier modes with the frequency of $0.01$\,$h$\,Mpc$^{-1}$.

As shown in Ref.~\cite{Taruya_Nishimichi_Jeong2018}, we have migrated the spurious aliasing contribution by introducing the high-$k$ cutoff. That is, we apply the sharp-$k$ filter of $k_{\rm cut,1}=1\,h$\,Mpc$^{-1}$ to the linear density fields, and then apply the same filter with $k_{\rm cut,2}=(4/3)\,h$\,Mpc$^{-1}$ to the higher-order density fields. 
%%%%%%%%%%%%%%%%%%%%%%%%%%%%%%%%%%%%%%%%%%%%%%%%%%%%%%%%%%%%%%%%%%%%%%%
\begin{figure*}[tb]
%\vspace*{-0.8cm}
\begin{center}
%\hspace*{0.5cm}
\includegraphics[width=13cm,angle=0]{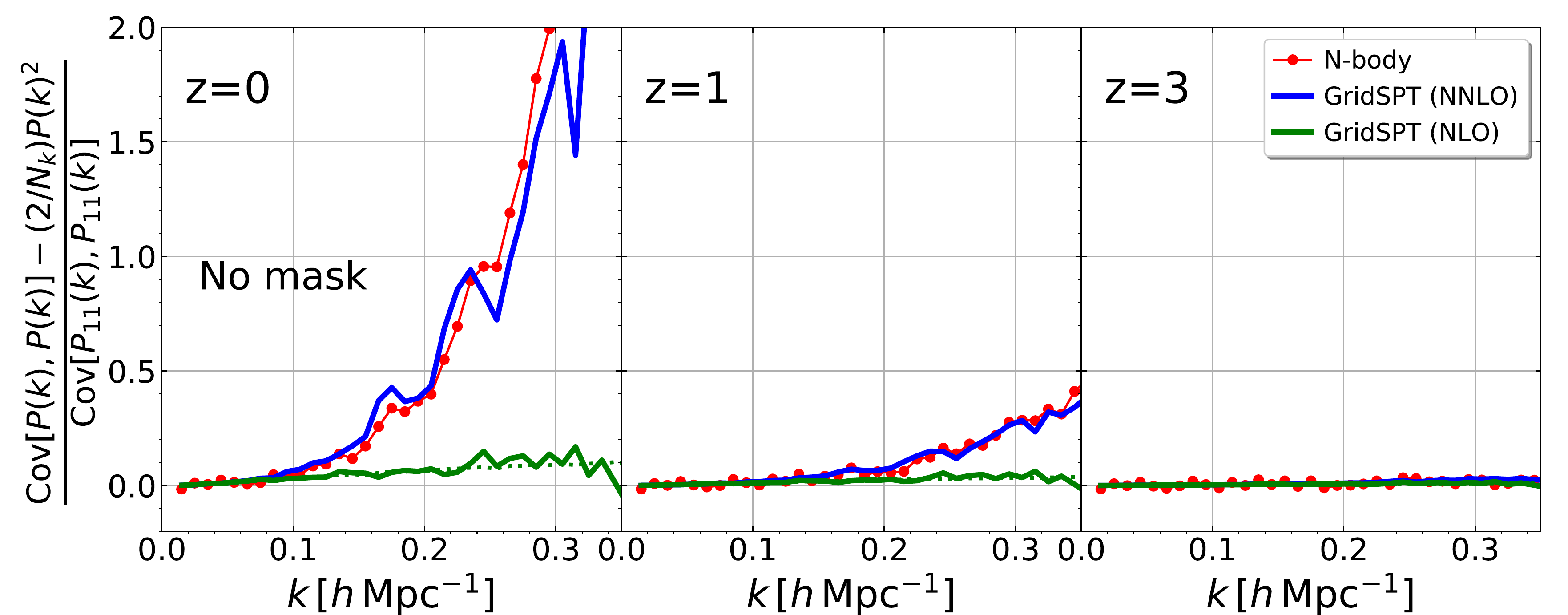}
\end{center}
\vspace*{-0.5cm}
\caption{Non-Gaussian contribution to the diagonal components of covariance matrices obtained from \gridspt\, and $N$-body simulations. The plotted results are $\{\cov[P(k_i,\,P(k)]-(2/N_k)P(k)^2\}/\cov[P_{11}(k),\,P_{11}(k)]$, and we show here the results at $z=0$ (left), $1$ (middle) and $3$ (right). The green and blue solid lines are respectively the NLO and NLLO results of the \gridspt, while the green dotted lines are the analytic SPT results including the tree-level trispectrum. The $N$-body results are shown in red filled circles.  
\label{fig:pkcov_diag_gridSPT_no_mask}
}
%\end{figure*}
%%%%%%%%%%%%%%%%%%%%%%%%%%%%%%%%%%%%%%%%%%%%%%%%%%%%%%%%%%%%%%%%%%%%%%%
%%%%%%%%%%%%%%%%%%%%%%%%%%%%%%%%%%%%%%%%%%%%%%%%%%%%%%%%%%%%%%%%%%%%%%%
%\begin{figure*}[tb]
%\vspace*{-0.8cm}
\begin{center}
%\hspace*{0.5cm}
\includegraphics[width=12cm,angle=0]{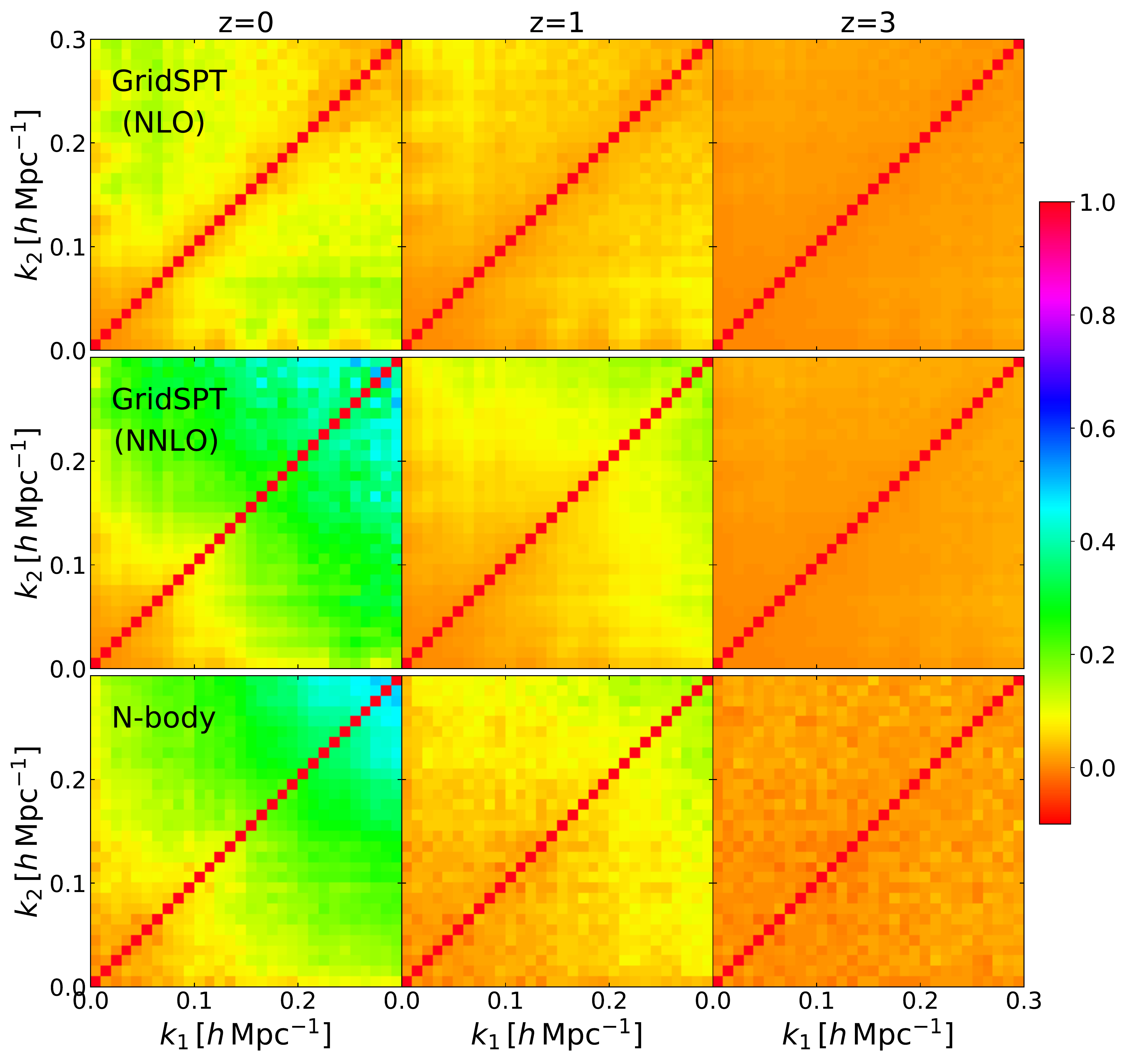}
\end{center}
\vspace*{-0.5cm}
\caption{Covariance matrices obtained from \gridspt\,(upper and middle) and $N$-body simulations (lower). The plotted results are the correlation coefficient matrix, $r(k_1,k_2)$, defined at Eq.~(\ref{eq:ratio_cov}), and we show here the results at $z=0$ (left), $1$ (middle) and $3$ (right).  For \gridspt\, results, the PT calculations of the covariance matrix at NLO and NNLO are respectively shown at upper and middle panels. 
\label{fig:pkcov_gridSPT_no_mask}
}
\end{figure*}
%%%%%%%%%%%%%%%%%%%%%%%%%%%%%%%%%%%%%%%%%%%%%%%%%%%%%%%%%%%%%%%%%%%%%%%

Given the grid-based density field in a cubic box, the power spectra are measured and the covariance are computed using many realizations in both cases with and without the survey window function. In Sec.~\ref{subsec:no_mask}, we compare the power spectrum covariance from \gridspt\, with that from the $N$-body simulation without the survey window function. As for the survey window function effect, for illustrative purpose, we consider the four simplified setups of survey masks shown in Fig.~\ref{fig:config_masks}, where the window function $W(\bfx)$ takes either $1$ or $0$, depending on the position inside the cubic box. Although these are rather idealistic setups far from reality, with the sharp cutoff at the boundary of the masks, their window functions exhibit a prominent feature in Fourier space, i.e., Bragg peak at high-$k$ for mosaic 2 and 4, and lack of large-scale modes for sphere 1 and 2. One would thus expect a significant impact on the off-diagonal part of the covariance matrix, and the setups in Fig.~\ref{fig:config_masks} therefore serve severe testing grounds for a robustness of our covariance estimation discussed in Sec.~\ref{sec:GridSPT_covariance}. We shall check it in detail in Sec.~\ref{subsec:mask}.

%We shall below check if the prescription given in Sec.~\ref{sec:GridSPT_covariance} provides a robust covariance estimation against there four survey window functions 

%%--%%--%%--%%--%%--%%--%%--%%--%%--%%--%%--%%--%%--%%--%%--%%--%%--%%
%%--%%--%%--%%--%%--%%--%%--%%--%%--%%--%%--%%--%%--%%--%%--%%--%%--%%
\subsection{Results without mask}
\label{subsec:no_mask}
%%--%%--%%--%%--%%--%%--%%--%%--%%--%%--%%--%%--%%--%%--%%--%%--%%--%%
%%--%%--%%--%%--%%--%%--%%--%%--%%--%%--%%--%%--%%--%%--%%--%%--%%--%%

Let us first present the results ignoring the survey window function. After examining the accuracy of the \gridspt~ calculation in terms of the power spectrum in Sec.~\ref{subsubsec:no_mask_pk}, the predicted covariance of \gridspt\, is presented up to the NNLO (fifth-order in perturbation theory), and is compared in detail with $N$-body simulations in Sec.~\ref{subsubsec:no_mask_covariance}. With the obtained power spectrum covariance, we have estimated the cumulative signal-to-noise ratio of nonlinear power spectrum in Sec.~\ref{subsubsec:no_mask_covariance_snr}.

%%--%%--%%--%%--%%--%%--%%--%%--%%--%%--%%--%%--%%--%%--%%--%%--%%--%%
\subsubsection{Power spectrum}
\label{subsubsec:no_mask_pk}
%%--%%--%%--%%--%%--%%--%%--%%--%%--%%--%%--%%--%%--%%--%%--%%--%%--%%

Fig.~\ref{fig:pk_gridSPT_no_mask} shows the power spectrum results at $z=0$ (left), $1$ (middle), and $3$ (right). Here, the results depicted as solid lines are the power spectra obtained from the \gridspt, averaging over $100,000$ realizations. The results from $N$-body simulations are also the averaged spectra, and we plot them with errorbars which indicate the standard deviation obtained from the diagonal of the measured covariance. In \gridspt~ results, three different colors represent the results at linear (black), one-loop (green, next-to-leading) and two-loop (blue, next-next-to-leading) order, which are constructed with the estimator of the SPT power spectrum at Eq.~(\ref{eq:estimator_Pab}) through
%%%%%%%%%%%%%%%%%%%%%%%%%%%%%%%%%%%%%%%%%%%%%%%%%%%%%%%%%%%%%%%%%%%%%%%
\begin{align}
 P(k)&= e^{2\eta}\,P_{\rm lin}(k)+ e^{4\eta}\,P_{\rm 1\mbox{-}loop}(k)+ e^{6\eta}\,P_{\rm 2\mbox{-}loop}(k)\,;
\nonumber
\\
& P_{\rm lin}(k) = P_{11}(k),
\label{eq:P_lin}
\\
& P_{\rm 1\mbox{-}loop}(k) = 2\,P_{13}(k) + P_{22}(k),
\label{eq:P_1loop}
\\
& P_{\rm 2\mbox{-}loop}(k) = 2\,P_{15}(k) + 2\,P_{24}(k) +P_{33}(k),
\label{eq:P_2loop}
\end{align}
%%%%%%%%%%%%%%%%%%%%%%%%%%%%%%%%%%%%%%%%%%%%%%%%%%%%%%%%%%%%%%%%%%%%%%%
where the quantities without hat imply the mean power spectra.

In Fig.~\ref{fig:pk_gridSPT_no_mask}, the discrepancies between the simulation and \gridspt~results are mostly ascribed to the impact of the nonlinear evolution that cannot be captured by the one- and two-loop corrections. While the qualitative trends of the discrepancies are similar to what have been seen in the literature (see e.g., Ref.~\cite{Nishimichi:2008ry,Taruya:2009ir}), the range of the agreement between the two-loop \gridspt~ and $N$-body results looks somewhat better and worse than expected at $z=0$ and $1$, respectively. We have checked that the measured power spectra from $N$-body simulations accurately agree well with predictions calibrated with high-resolution $N$-body simulations based on the response function approach \cite{Nishimichi_etal2017}. Thus, the trends seen at $z=0$ and $z=1$ are presumably due to the imperfect elimination of the aliasing effect in \gridspt\, calculations with our choice of the cutoff scale (see Sec.~\ref{subsec:setup}). As it has been discussed in detail in Ref.~\cite{Taruya_Nishimichi_Jeong2018}, the aliasing effect can systematically change the power spectrum, and the effect dominantly comes from the higher-loop corrections. Thus, its impact can be significant at lower redshifts. Since the two-loop correction of the power spectrum is rather sensitive to the high-$k$ cutoff, a further study is required for choosing the optimal cutoff scale. Here, however, we simply adopt the same cutoff scales as used in the previous paper (Ref.~\cite{Taruya_Nishimichi_Jeong2018}), because as we shall see below, this does not affect the covariance calculation so much. In fact, the \gridspt\, covariance shows a reasonable behavior which quantitatively explains measured results from $N$-body simulations. 

%due to \gridspt, in particular, originating from our choice of the cutoff scales to mitigate the aliasing effect (see Sec.~\ref{subsec:setup}). As it has been discussed in detail in Ref.~\cite{Taruya_Nishimichi_Jeong2018}, 

%%--%%--%%--%%--%%--%%--%%--%%--%%--%%--%%--%%--%%--%%--%%--%%--%%--%%
\subsubsection{Covariance matrix}
\label{subsubsec:no_mask_covariance}
%%--%%--%%--%%--%%--%%--%%--%%--%%--%%--%%--%%--%%--%%--%%--%%--%%--%%

Let us now turn to the results of the covariance matrix, focusing on their non-Gaussian contributions. 

Fig.~\ref{fig:pkcov_diag_gridSPT_no_mask} shows the diagonal part of the covariance matrix normalized by that of the linear-order power spectrum, also subtracting the Gaussian contribution, i.e., 
%%%%%%%%%%%%%%%%%%%%%%%%%%%%%%%%%%%%%%%%%%%%%%%%%%%%%%%%%%%%%%%%%%%%%%%
\begin{align}
\frac{\cov[P(k),P(k)]-(2/N_k)P(k)^2}{\cov[P_{11}(k),P_{11}(k)]}.  
\label{eq:ratio_cov_diag}
\end{align}
%%%%%%%%%%%%%%%%%%%%%%%%%%%%%%%%%%%%%%%%%%%%%%%%%%%%%%%%%%%%%%%%%%%%%%%
Note that when subtracting the Gaussian contribution, we used the power spectrum averaging over realizations. For \gridspt, the expressions summarized at Eqs.~(\ref{eq:covSPT_Lin_2nd})--(\ref{eq:covSPT_NNLO_2nd}) are used to identify the Gaussian contributions at each order, and the terms involving the Kronecker delta in their expressions are subtracted from Eqs.~(\ref{eq:PT_covariance_Lin})--(\ref{eq:PT_covariance_NNLO}). The green and blue solid lines are respectively the \gridspt\, results at NLO and NNLO, while the filled red circles with lines are the measured covariance from $N$-body simulations. At $z=3$, the non-Gaussian contribution to the covariance is negligibly small, and all the results coincide with each other. As decreasing the redshifts, however, we observe the development of significant amount of the non-Gaussianity. While consistently reproducing the analytical SPT results depicted as green dotted lines, the \gridspt\, results at NLO significantly underpredict the simulation results. Adding the higher-order corrections, the \gridspt\, covariance at NNLO reproduces quantitatively the N-body simulation results at $z=1$, and even at $z=0$, it gives a reasonable agreement.

\begin{figure*}[tb]
\vspace*{-0.5cm}
\begin{center}
%\hspace*{0.5cm}
\includegraphics[width=15cm,angle=0]{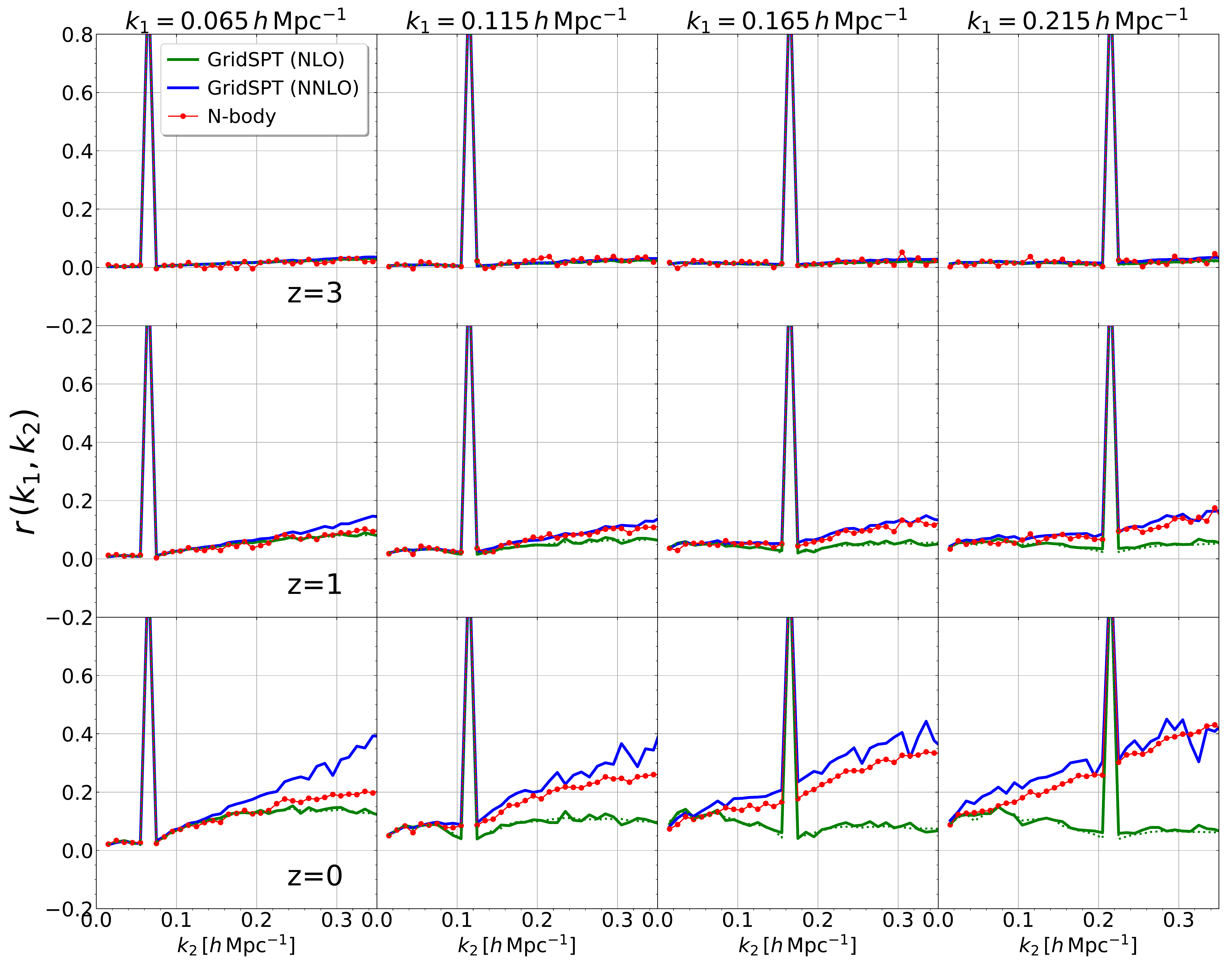}
\end{center}
\vspace*{-0.5cm}
\caption{Power spectrum covariance at $z=3$ (upper), $1$ (middle) and $0$ (lower) obtained from {\tt GridSPT} calculations, specifically fixing $k_1$ to $0.065$, $0.115$, $0.165$, and $0.215\,h$\,Mpc$^{-1}$ (from left to bottom). Red filled circles are the $N$-body results. Green and blue solid lines are the \gridspt\, covariance at NLO and NNLO, respectively including the tree-level and one-loop trispectrum. For comparison, the analytic SPT results including the tree-level trispectrum are also shown in green dotted lines. 
\label{fig:pkcov_gridSPT_k1_5_10_15_20bins_no_mask}
}
%\end{figure*}
%%%%%%%%%%%%%%%%%%%%%%%%%%%%%%%%%%%%%%%%%%%%%%%%%%%%%%%%%%%%%%%%%%%%%%%
%%%%%%%%%%%%%%%%%%%%%%%%%%%%%%%%%%%%%%%%%%%%%%%%%%%%%%%%%%%%%%%%%%%%%%%
%\begin{figure*}[tb]
%\vspace*{-0.8cm}
\begin{center}
%\hspace*{0.5cm}
\includegraphics[width=15cm,angle=0]{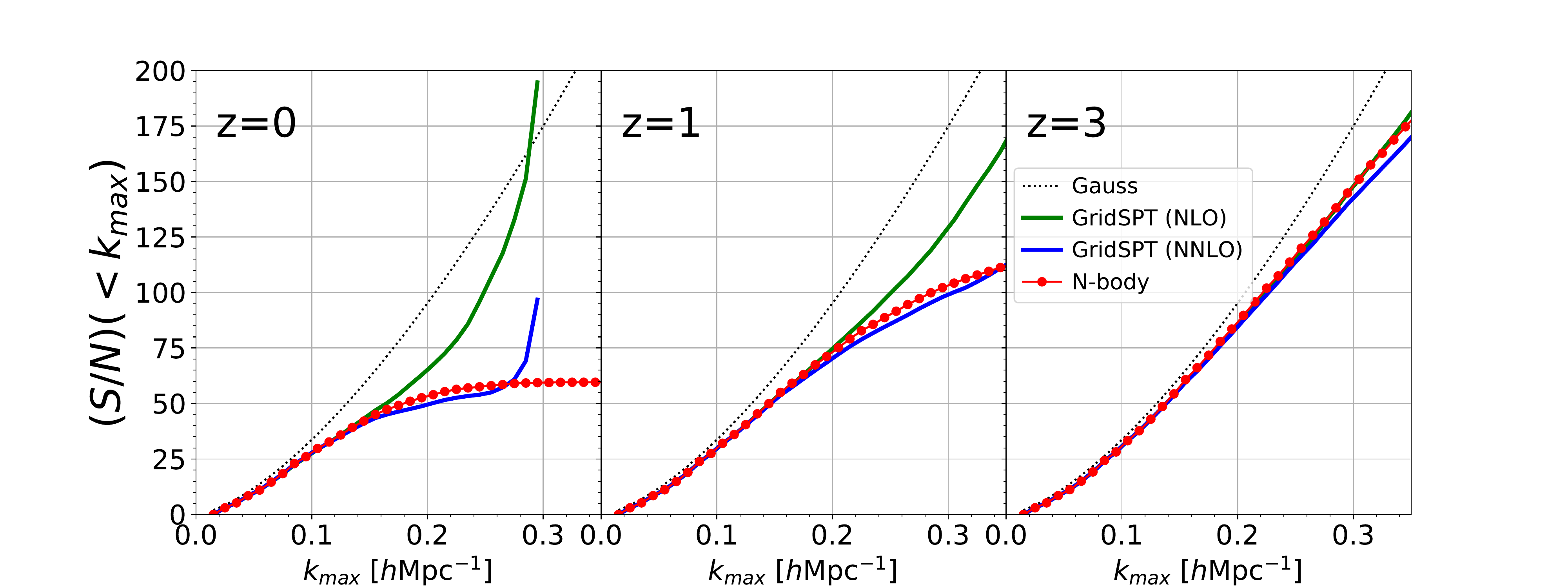}
\end{center}
\vspace*{-0.5cm}
\caption{Signal-to-noise ratios of the power spectrum at $z=0$ (left), $1$ (middle) and $3$ (right), with the off-diagonal components of covariance matrix computed from {\tt GridSPT} calculations at NLO (green) and NNLO (blue). Filled red symbols connected with solid line are all computed with $N$-body simulations. The dotted lines are the expected signal-to-noise ratios in the Gaussian limit, which are independent of redshift. 
\label{fig:snr_gridSPT_no_mask}
}
\end{figure*}

Next, we focus on the off-diagonal components. Fig.~\ref{fig:pkcov_gridSPT_no_mask} shows the structure of the off-diagonal components measured at $z=0$ (left), $1$ (middle), and $3$ (right). The results of the \gridspt\, calculations at NLO and NNLO are plotted in upper two panels, and these are compared with the $N$-body results, shown in the bottom panel. Further, in Fig.~\ref{fig:pkcov_gridSPT_k1_5_10_15_20bins_no_mask},  the results at four selected wavenumbers $k_1$, as indicated at the top of each panel, are particularly shown, plotted as a function of $k_2$. In all cases, we show here the correlation coefficient matrix defined by (e.g., \cite{Takahashi2009,Blot2015})
%%%%%%%%%%%%%%%%%%%%%%%%%%%%%%%%%%%%%%%%%%%%%%%%%%%%%%%%%%%%%%%%%%%%%%%
\begin{align}
&r(k_1,k_2)
\nonumber
\\
& \equiv \frac{\cov[P(k_1),P(k_2)]}{\sqrt{\cov[P_{\rm sim}(k_1),P_{\rm sim}(k_1)]\, \cov[P_{\rm sim}(k_2),P_{\rm sim}(k_2)]} },
\label{eq:ratio_cov}
\end{align}
%%%%%%%%%%%%%%%%%%%%%%%%%%%%%%%%%%%%%%%%%%%%%%%%%%%%%%%%%%%%%%%%%%%%%%%
where $P_{\rm sim}$ is the measured power spectrum in $N$-body simulations. 
In plotting the \gridspt\, results in Figs.~\ref{fig:pkcov_gridSPT_no_mask} and \ref{fig:pkcov_gridSPT_k1_5_10_15_20bins_no_mask}, just for illustrative purpose to compare the three results in a common ground, we divide the covariance by the diagonal component of the $N$-body results. Therefore, the diagonal components ($k_1=k_2$) reads unity only for the $N$-body cases. Substituting the \gridspt~results into the numerator, the above quantity does not ensure the Schwarz inequality in general, and it can eventually exceed 1 or fall off below $-1$ for the off-diagonal components.

%%%%%%%%%%%%%%%%%%%%%%%%%%%%%%%%%%%%%%%%%%%%%%%%%%%%%%%%%%%%%%%%%%%%%%%
%%%%%%%%%%%%%%%%%%%%%%%%%%%%%%%%%%%%%%%%%%%%%%%%%%%%%%%%%%%%%%%%%%%%%%%

On the large scales ($k<0.3\,h\,$Mpc$^{-1}$) that we show in Figs.~\ref{fig:pkcov_gridSPT_no_mask} and \ref{fig:pkcov_gridSPT_k1_5_10_15_20bins_no_mask}, off-diagonal components of the covariance matrix typically have $r(k_i,k_j)\lesssim0.3-0.4$ at $z=0$ (e.g., Refs.~\cite{Takahashi2009,Blot2015}), and except the vicinity of the diagonal components, it is a monotonically increasing function of $k_1$ and $k_2$. Looking at Fig.~\ref{fig:pkcov_gridSPT_no_mask}, the NNLO results of \gridspt, which includes the one-loop trispectrum contribution [see Eq.~(\ref{eq:covSPT_NNLO_2nd})], reproduce well the trends seen in the simulations at all three redshifts. Thanks to a large (100,000) number of realizations, all the \gridspt\, results are less noisy than the $N$-body covariance measured from $10,000$ simulations. 

Including the tree-level trispectrum, the NLO results of \gridspt\, also provides a reasonable match at $z=3$, but the differences are manifest at lower redshifts; the NLO results underestimate the N-body results. This is indeed clearly seen in Fig.~\ref{fig:pkcov_gridSPT_k1_5_10_15_20bins_no_mask}, where the NLO results, depicted as green lines, are found typically to give $r(k_i,k_j)\sim0.1$ at $z=0$. Note that similarly to the diagonal part, we see a nice agreement between the \gridspt\, (solid) and analytic SPT (dotted, using tree-level trispectrum) calculations, ensuring a correct implementation of the grid-based calculation of the power-spectrum covariance.
%%%%%%%%%%%%%%%%%%%%%%%%%%%%%%%%%%%%%%%%%%%%%%%%%%%%%%%%%%%%%%%%%%%%%%%
\begin{figure*}[tb]
%\vspace*{-0.8cm}
\begin{center}
%\hspace*{0.5cm}
\includegraphics[width=15cm,angle=0]{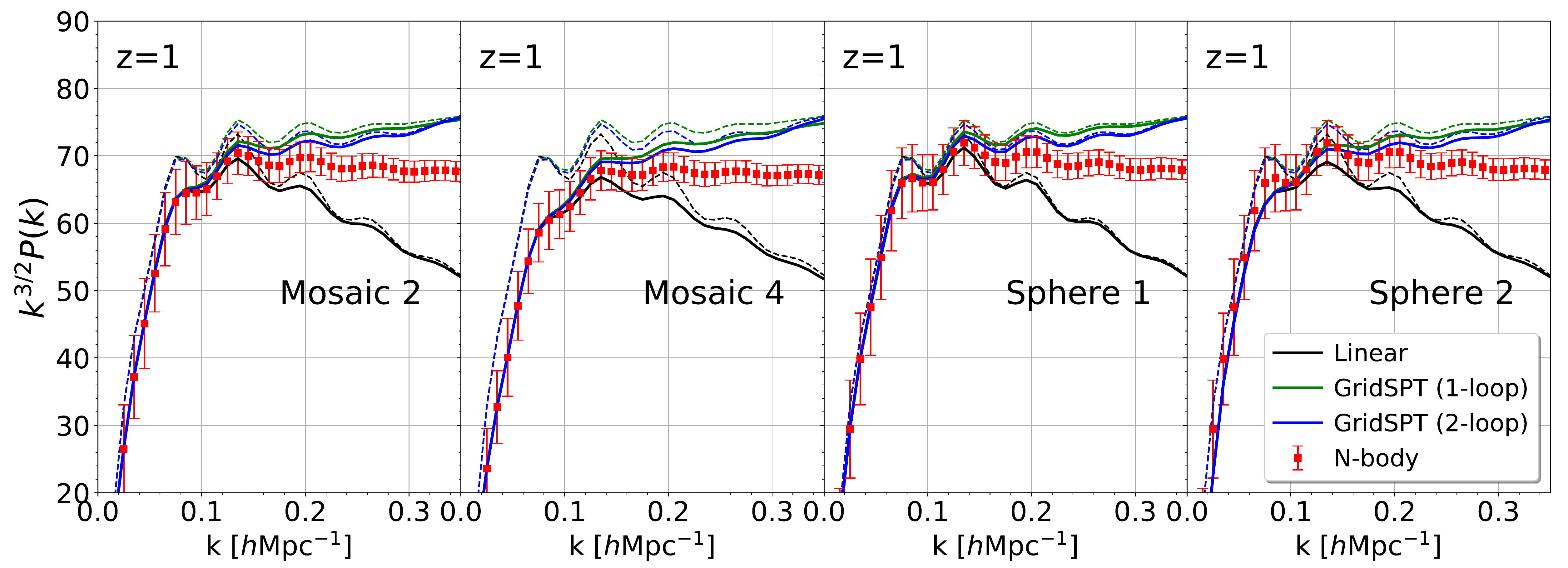}
\end{center}
\vspace*{-0.5cm}
\caption{Power spectra at $z=1$, taking the survey masks into account. Meanings of the colored lines and symbols are the same as in Fig.~\ref{fig:pk_gridSPT_no_mask}. For reference, \gridspt\,results without taking masks are also shown in dotted lines. 
\label{fig:pk_gridSPT_z1_with_mask}
}
\end{figure*}
%%%%%%%%%%%%%%%%%%%%%%%%%%%%%%%%%%%%%%%%%%%%%%%%%%%%%%%%%%%%%%%%%%%%%%%

Adding the higher-order contributions, the \gridspt\, covariance at NNLO takes a larger value than the NLO results, and it closely matches the simulation results at $z=1$. A closer look at $z=0$, however, reveals that even the NNLO results tend to overpredict the simulations, particularly when either $k_1$ or $k_2$ are larger than $0.1\,h$\,Mpc$^{-1}$. While the level of agreement between \gridspt\, and $N$-body simulations is qualitatively similar to what we saw in the power spectrum, the deviation in the covariance starts from smaller wavenumbers (larger scales). This indicates that the \gridspt\, covariance at NNLO receives more impact from the trispectrum than the power spectrum at large scales. This might be potentially ascribed to the UV-sensitive behaviors of the single-stream PT treatment, as it has been recently advocated (e.g., \cite{Blas:2013aba,Bernardeau:2012ux}), and their impact may be more significant for the higher-order statistics. The effective-field-theory treatment in Ref.~\cite{Bertolini2016} has hinted the signatures of the UV-sensitivity in the covariance calculations from PT.

For completeness, in Appendix~\ref{sec:covariance_tree_1loop}, we break down the non-Gaussian contributions to the power spectrum covariance, and show individual PT term as well as the partial summations. In the figures in the appendix, one can find the usual cancellations appearing in PT calculations.

%%%%%%%%%%%%%%%%%%%%%%%%%%%%%%%%%%%%%%%%%%%%%%%%%%%%%%%%%%%%%%%%%%%%%%%
\begin{figure*}[tb]
%\vspace*{-0.8cm}
\begin{center}
%\hspace*{0.5cm}
\includegraphics[width=13cm,angle=0]{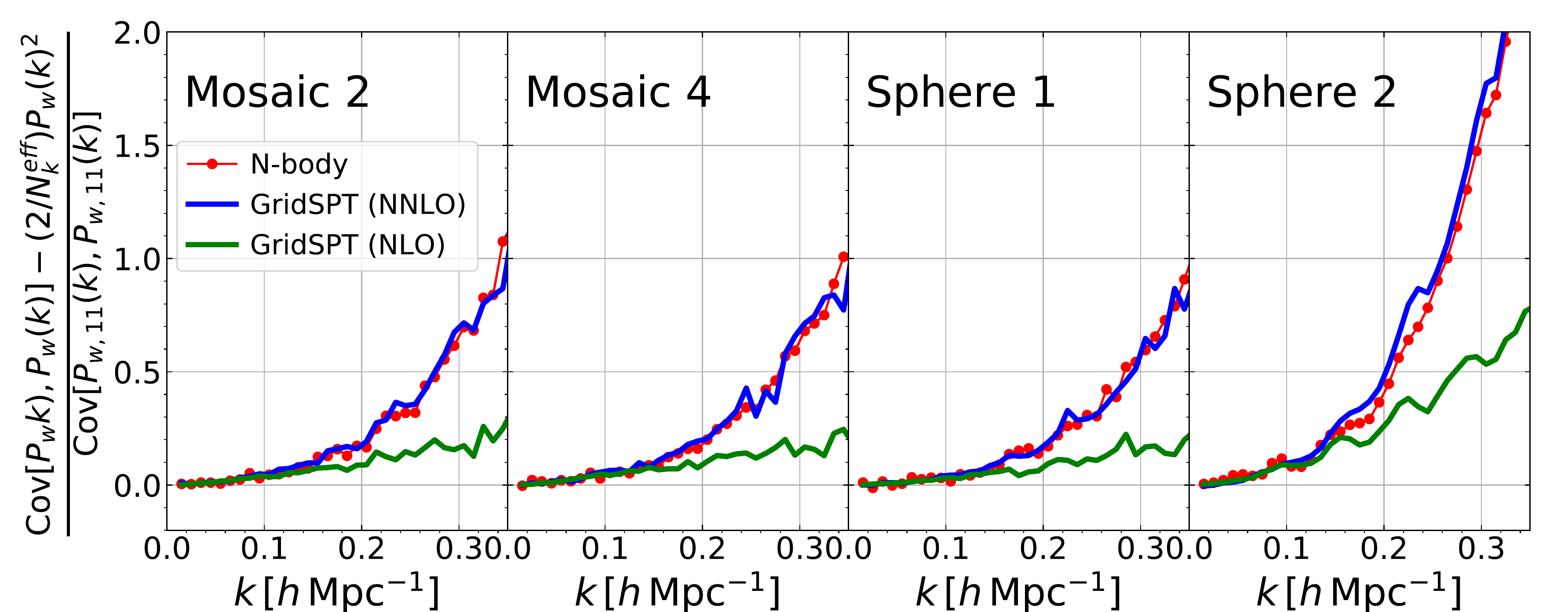}
\end{center}
\vspace*{-0.5cm}
\caption{Diagonal components of the covariance matrices taking the survey masks into account. The plotted results are similar to those in Fig.~\ref{fig:pkcov_diag_gridSPT_no_mask}, but the results at $z=1$ are shown, also with Eq.~(\ref{eq:ratio_cov_diag}) modified according to the presence of survey masks (see text). The line types are the same as in Fig.~\ref{fig:pkcov_diag_gridSPT_no_mask}. 
\label{fig:pkcov_diag_gridSPT_z1_with_mask}
}
%\end{figure*}
%%%%%%%%%%%%%%%%%%%%%%%%%%%%%%%%%%%%%%%%%%%%%%%%%%%%%%%%%%%%%%%%%%%%%%%
%%%%%%%%%%%%%%%%%%%%%%%%%%%%%%%%%%%%%%%%%%%%%%%%%%%%%%%%%%%%%%%%%%%%%%%
%\begin{figure*}[tb]
%\vspace*{-0.8cm}
\begin{center}
%\hspace*{0.5cm}
\includegraphics[width=15cm,angle=0]{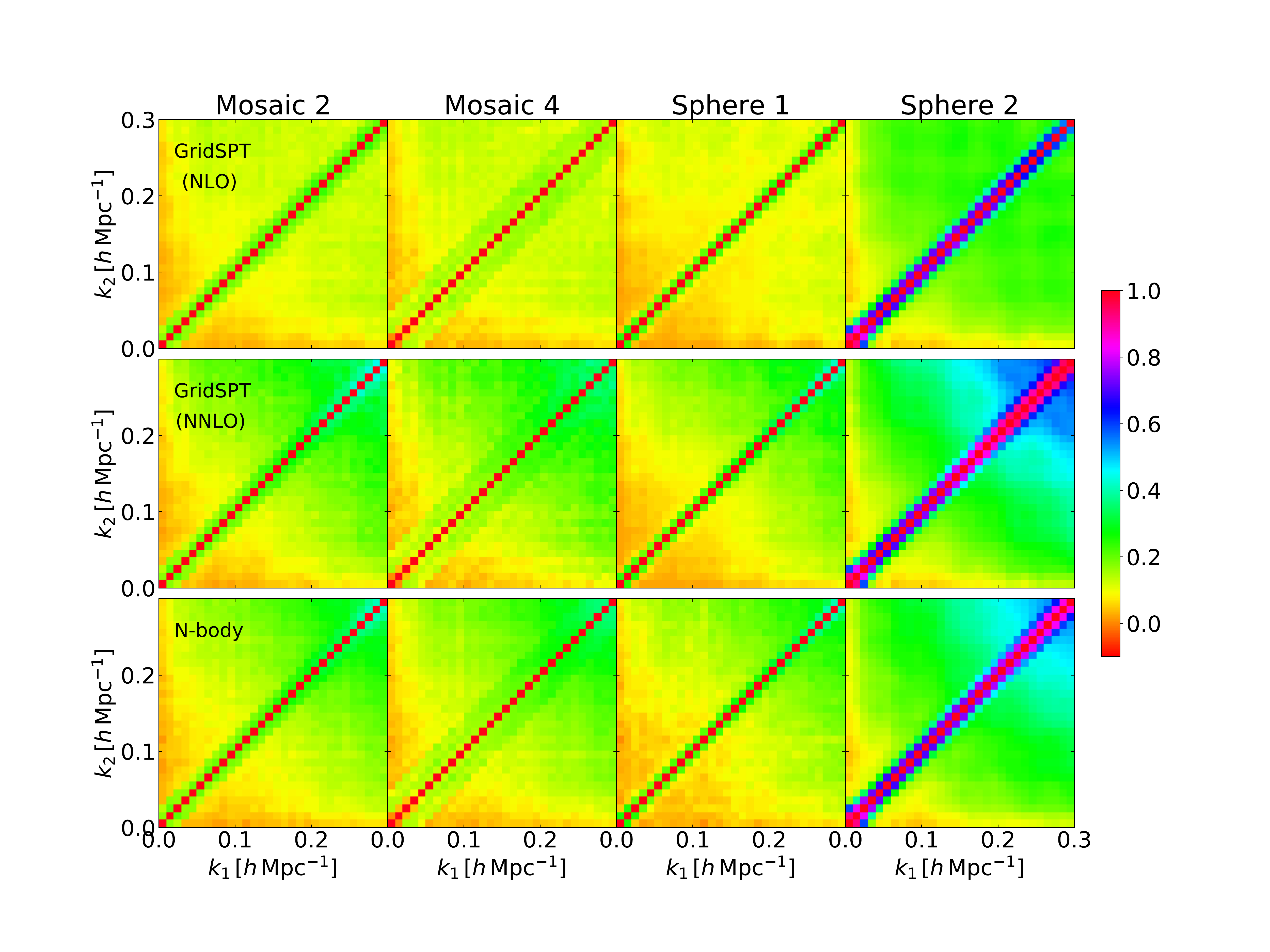}
\end{center}
\vspace*{-0.5cm}
\caption{Covariance matrices at $z=1$, taking into account the survey masks shown in Fig.~\ref{fig:config_masks}. The plotted results are the correlation coefficient matrix, $r(k_1,k_2)$, defined at Eq.~(\ref{eq:ratio_cov}). Upper and middle panels are respectively the \gridspt\, results at NLO and NNLO. Bottom panels are the results measured from $N$-body simulations. 
\label{fig:pkcov_gridSPT_z1_with_mask}
}
\end{figure*}
%%%%%%%%%%%%%%%%%%%%%%%%%%%%%%%%%%%%%%%%%%%%%%%%%%%%%%%%%%%%%%%%%%%%%%%

%%--%%--%%--%%--%%--%%--%%--%%--%%--%%--%%--%%--%%--%%--%%--%%--%%--%%
\subsubsection{Signal-to-noise ratio}
\label{subsubsec:no_mask_covariance_snr}
%%--%%--%%--%%--%%--%%--%%--%%--%%--%%--%%--%%--%%--%%--%%--%%--%%--%%

To facilitate the comparison among the power spectrum covariances from \gridspt\, NLO, \gridspt\, NNLO and $N$-body simulation results that we have discussed in Sec.~\ref{subsubsec:no_mask_covariance}, we estimated the cumulative signal-to-noise ratio, $(S/N)$, of the power spectrum defined as  
%%%%%%%%%%%%%%%%%%%%%%%%%%%%%%%%%%%%%%%%%%%%%%%%%%%%%%%%%%%%%%%%%%%%%%%
\begin{align}
& \Bigl(\frac{S}{N}\Bigr)^2(<k_{\rm max})
\nonumber
\\
&\equiv \sum_{k_i,k_j
\le k_{\rm max}}\,P_{\rm sim}(k_i)\,\{\cov[P(k_i),P(k_j)]\}^{-1}\,P_{\rm sim}(k_j),
\label{eq:def_snr}
\end{align}
%%%%%%%%%%%%%%%%%%%%%%%%%%%%%%%%%%%%%%%%%%%%%%%%%%%%%%%%%%%%%%%%%%%%%%%
which depends on the entirety of the power spectrum covariance.

Fig.~\ref{fig:snr_gridSPT_no_mask} shows the cumulative signal-to-noise ratio as a function of the maximum wavenumber $k_{\rm max}$. Remarkably, the agreement between the \gridspt\, covariance at NNLO (blue) and the N-body result is excellent, indicating that a discrepancy found at the off-diagonal part does not affect much to the total signal-to-noise ratio. By contrast, on small scales, the covariance at NLO (green) significantly overestimate the signal-to-noise ratio at lower redshifts.

At $z=0$, the signal-to-noise ratio from the NLO \gridspt\, significantly increases and exhibits a divergent behavior at $k_{\rm max}\sim0.3\,h$\,Mpc$^{-1}$, eventually exceeding the expected $(S/N)$ in the Gaussian limit (dotted). The result from the \gridspt\, covariance at NNLO also shows a similar behavior. The divergence of the predicted $(S/N)$ was also seen in Ref.~\cite{Takahashi2009} (their Fig.~7). We have checked from the analytical PT calculation of NLO, that these behaviors appear when the covariance matrix becomes singular and non-invertible. Note, however, that such a divergence basically appears at the scale where the PT prediction of the power spectrum fails to reproduce the $N$-body results (see Fig.~\ref{fig:pk_gridSPT_no_mask}). Moreover, for more practical situations, when shot-noise contribution dominates the power spectrum covariance on small scales, such divergence does not appear (e.g., \cite{Digvijay_Scoccimarro2019,Sugiyama_etal2019}). Therefore, within the valid range of SPT calculations, the divergence can be ignored, and the estimation based on the NNLO calculations provides a good description for the signal-to-noise ratio at all redshifts.

%%%%%%%%%%%%%%%%%%%%%%%%%%%%%%%%%%%%%%%%%%%%%%%%%%%%%%%%%%%%%%%%%%%%%%%
\begin{figure*}[tb]
%\vspace*{-0.8cm}
\begin{center}
%\hspace*{0.5cm}
\includegraphics[width=15cm,angle=0]{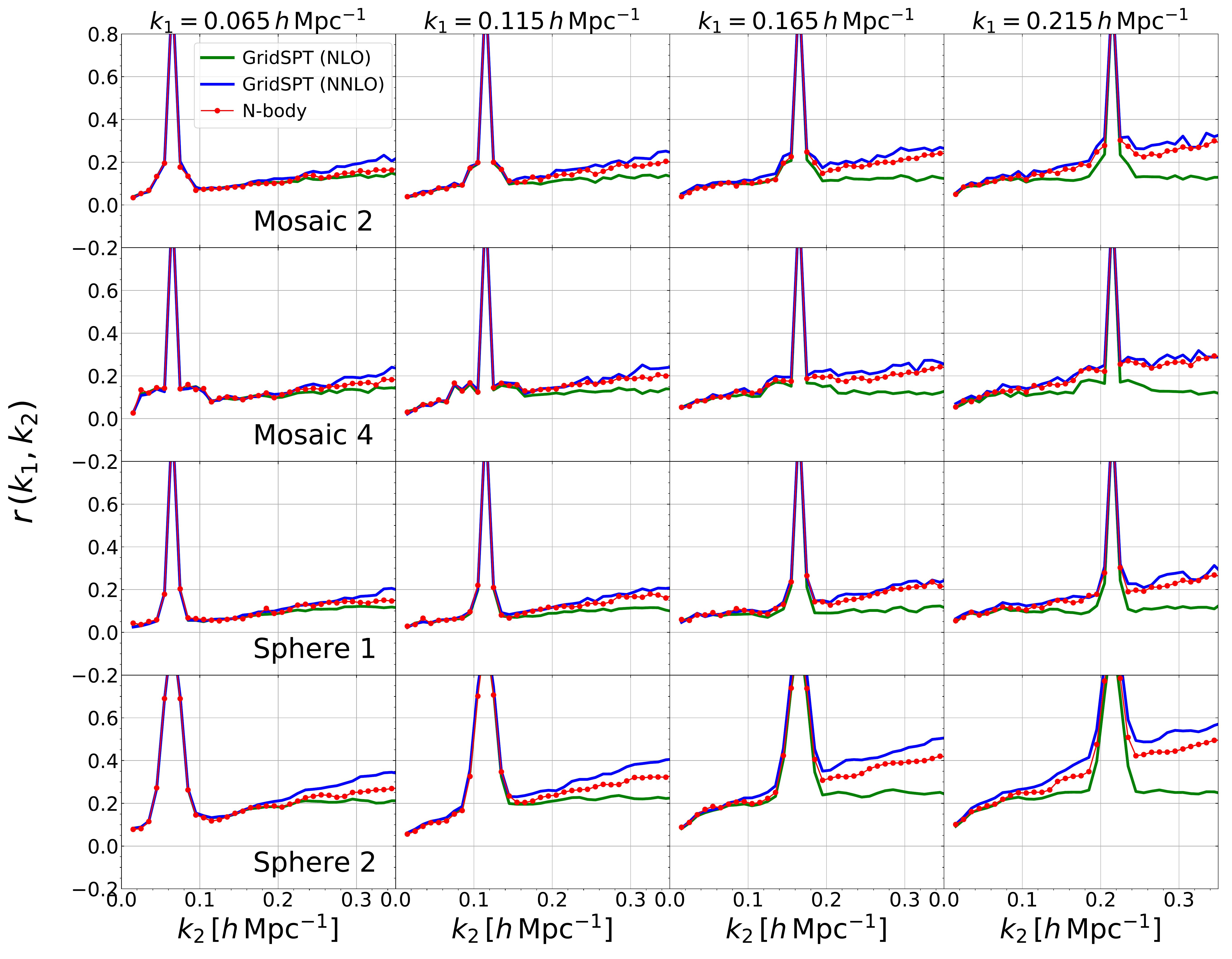}
\end{center}
\vspace*{-0.5cm}
\caption{Covariance matrices at $z=1$, taking the survey masks shown in Fig.~\ref{fig:config_masks} into account. Specifically fixing $k_1$ to $0.065$, $0.115$, $0.165$, and $0.215\,h$\,Mpc$^{-1}$, the correlation coefficient matrix, $r(k_i,k_j)$, are shown from left to bottom panels, plotted as a function of $k_2$. Meaning of colored lines and symbols are the same as in Fig.~\ref{fig:pkcov_gridSPT_k1_5_10_15_20bins_no_mask}. 
\label{fig:pkcov_gridSPT_z1_k1_5_10_15_20bins_with_mask}
}
%\end{figure*}
%%%%%%%%%%%%%%%%%%%%%%%%%%%%%%%%%%%%%%%%%%%%%%%%%%%%%%%%%%%%%%%%%%%%%%%
%%%%%%%%%%%%%%%%%%%%%%%%%%%%%%%%%%%%%%%%%%%%%%%%%%%%%%%%%%%%%%%%%%%%%%%
%\begin{figure*}[tb]
%\vspace*{-0.8cm}
\begin{center}
%\hspace*{0.5cm}
\includegraphics[width=15cm,angle=0]{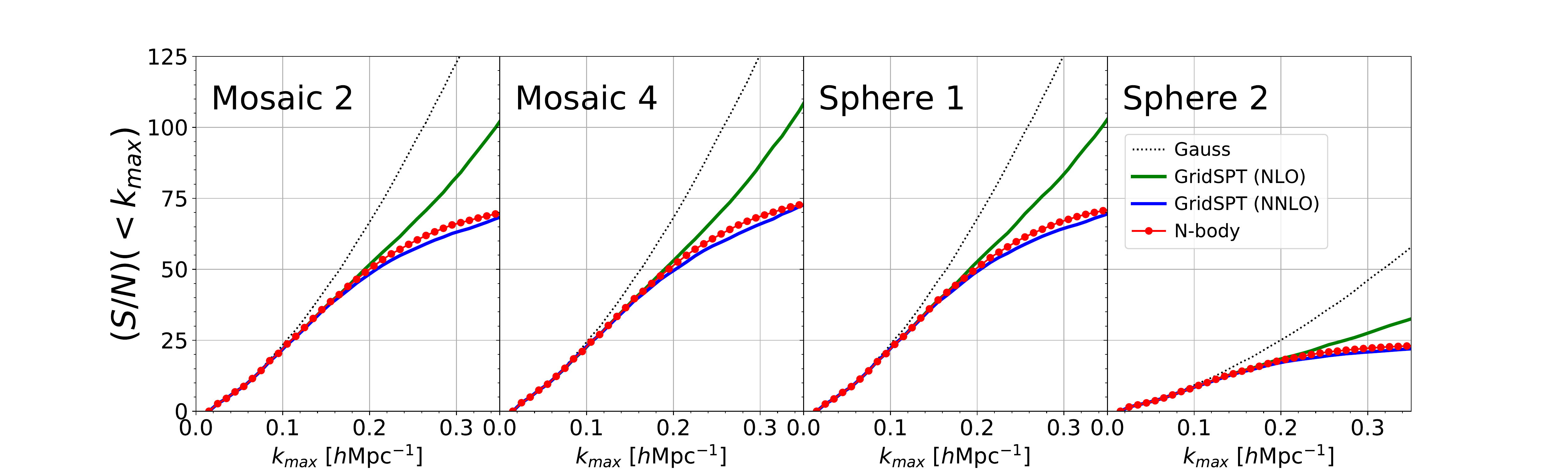}
\end{center}
\vspace*{-0.5cm}
\caption{Signal-to-noise ratios of power spectrum at $z=1$, taking the survey masks into account. In each panel, the results with different survey masks are shown. Meaning of line types and symbols are the same as those shown in Fig.~\ref{fig:snr_gridSPT_no_mask}. Note that the signal-to-noise ratios in the Gaussian limit, depicted as dotted lines, are estimated using the covariance measured from randomly generated linear density fields convolving the survey masksi.e., $\cov[P_{\rm w,11}(k_i),\,P_{\rm w,11}(k_j)]$. 
\label{fig:snr_gridSPT_z1_with_mask}
}
\end{figure*}
%%%%%%%%%%%%%%%%%%%%%%%%%%%%%%%%%%%%%%%%%%%%%%%%%%%%%%%%%%%%%%%%%%%%%%%

%%--%%--%%--%%--%%--%%--%%--%%--%%--%%--%%--%%--%%--%%--%%--%%--%%--%%
%%--%%--%%--%%--%%--%%--%%--%%--%%--%%--%%--%%--%%--%%--%%--%%--%%--%%
\subsection{Results with survey masks}
\label{subsec:mask}
%%--%%--%%--%%--%%--%%--%%--%%--%%--%%--%%--%%--%%--%%--%%--%%--%%--%%
%%--%%--%%--%%--%%--%%--%%--%%--%%--%%--%%--%%--%%--%%--%%--%%--%%--%%

Having confirmed that the \gridspt\, provides a robust way to estimate the power spectrum covariance, let us consider the cases with survey window. Below we shall specifically present the results at $z=1$. 

When the survey window function is defined as the survey mask ($W(\bfx)=0$ inside mask, $1$ outside of mask) and the mean density stays constant over the survey volume, the overall amplitude of the measured power spectra is suppressed by the ratio between the total volumes and the masked volume, $V_{\rm w}$ in Eq.~(\ref{eq:def_VW}). In order to make the results comparable to the one presented in Sec.~\ref{subsec:no_mask}, without survey masks, we multiply all the measured power spectra by $(V/V_{\rm w})$, and hence the covariance matrix by $(V/V_{\rm w})^2$.

%%--%%--%%--%%--%%--%%--%%--%%--%%--%%--%%--%%--%%--%%--%%--%%--%%--%%
\subsubsection{Power spectrum}
\label{subsubsec:masked_pk}
%%--%%--%%--%%--%%--%%--%%--%%--%%--%%--%%--%%--%%--%%--%%--%%--%%--%%

Fig.~\ref{fig:pk_gridSPT_z1_with_mask} shows the power spectra obtained from the density fields with the four survey masks that we show in Fig.~\ref{fig:config_masks}. Here, for reference, we also plot the results without the survey mask (Fig.~\ref{fig:pk_gridSPT_no_mask}) as short-dashed lines with the respective colors.

When the survey masks are considered, the shape of the power spectrum is deformed and the resultant amplitude is also changed. The BAO features are also modulated, and tend to be smeared even at linear scales. With the treatment described in Sec.~\ref{subsec:survey_window}, the \gridspt\, results quantitatively describe the $N$-body trends at large scales, although, similarly to what we saw in Sec.~\ref{subsec:no_mask}, the agreement between two-loop \gridspt\, and $N$-body results is not as good as one usually expected. Still, however, the Fourier-wavenumber range that the \gridspt\, prescription is valid remains almost unchanged irrespective of the survey masks.

%%--%%--%%--%%--%%--%%--%%--%%--%%--%%--%%--%%--%%--%%--%%--%%--%%--%%
\subsubsection{Covariance matrix}
\label{subsubsec:masked_pkcov}
%%--%%--%%--%%--%%--%%--%%--%%--%%--%%--%%--%%--%%--%%--%%--%%--%%--%%

Fig.~\ref{fig:pkcov_diag_gridSPT_z1_with_mask} shows the diagonal part of the covariance matrix at $z=1$. Similarly to what has been done in Fig.~\ref{fig:pkcov_diag_gridSPT_no_mask}, we subtract the contributions that correspond to the disconnected covariance in the absence of survey masks, and the results are normalized by the linear-order covariance. That is, the plotted results are Eq.~(\ref{eq:ratio_cov_diag}), but the power spectra $P(k)$ and $P_{11}(k)$ are replaced with those measured from the masked density fields, i.e., $P_{\rm w}(k)$ and $P_{\rm w,11}(k)$. Also, to account for the different weight for the convolved window functions between the power spectrum and covariance, the factor $2/N_k$ has to be properly replaced with $2/N_k^{\rm eff}$ with $N_k^{\rm eff}$ being the effective number of modes (e.g., \cite{dePutter_etal2012,Li_etal2019,Digvijay_Scoccimarro2019}). Here, to estimate this, we use the linear-order \gridspt\, results, and set $2/N_k^{\rm eff}$ to $\cov[P_{\rm w,11}(k),P_{\rm w,11}(k)]/\{P_{\rm w,11}(k)\}^2$ in both simulations and \gridspt\footnote{We have checked that in the absence of survey masks, this treatment accurately matches well with the expected result.}.

In comparison with Fig.~\ref{fig:pkcov_diag_gridSPT_no_mask}, the results in Fig.~\ref{fig:pkcov_diag_gridSPT_z1_with_mask} clearly show that the survey mask alters the mode-coupling structure and gives an impact on the diagonal part of the covariance matrix. The impact gets larger as increasing the wavenumber, and at $k\gtrsim0.3\,h$\,Mpc$^{-1}$, it amplifies the diagonal part of the connected covariance\footnote{Strictly speaking, adopting the density estimator at Eq.~(\ref{eq:estimator_delta_with_mask}), which ensures the vanishing local mean, the plotted results do not precisely correspond to the connected covariance originated from the trispectrum of the true density fields, but partly include the Gaussian contributions.} by more than a factor of $2$. Nevertheless, the \gridspt\, covariance explains these trends, and, in particular, the NNLO results reproduce the simulations quantitatively well, for all four types of the survey masks that we consider here.

Fig.~\ref{fig:pkcov_gridSPT_z1_with_mask} shows the off-diagonal part of 
the covariance matrices, and compares the \gridspt\, (top and middle) with $N$-body (bottom) results. In Fig.~\ref{fig:pkcov_gridSPT_z1_k1_5_10_15_20bins_with_mask}, for specific wavenumbers at $k_1=0.065$, $0.115$, $0.165$ and $0.215\,h$\,Mpc$^{-1}$, the off-diagonal covariance is plotted as a function of $k_2$. Note again that all the plotted results are the correlation coefficient matrix, $r(k_1,k_2)$ [Eq.~(\ref{eq:ratio_cov})], and in plotting the \gridspt\, results, the diagonal part of the covariance is replaced with the $N$-body results.

Compared to the case without survey masks, the off-diagonal components are more developed even at $z=1$, and the amplitude of $r(k_1,k_2)$ gets larger. Also, there appear characteristic structures near the diagonal part. These are solely due to the mode coupling through the survey window function. Although a detailed covariance structure depends on the properties of the survey window function, the survey mask of sphere 2 gives the largest impact on the resultant covariance among those we consider, and the amplitude of the correlation coefficient matrix is lifted up at both large and small scales. This is presumably due to the super-survey modes inside the cubic box, whose wavelength exceed the survey region.

Overall, including the NNLO contributions, the \gridspt\, covariance reproduces the $N$-body result very well, especially at $k_{1,2}\lesssim0.2\,h$\,Mpc$^{-1}$. Note that in the presence of survey masks, the Gaussian contributions, which are included in the \gridspt\, covariance at each order,  now play a very important role to describe the covariance structure near the diagonal components. Despite the survey window with a sharp contrast (having only $0$ or $1$), the resultant \gridspt\, covariance mostly accounts for the trends seen in the $N$-body results. A closer look at $k_{1,2}\gtrsim0.2\,h$\,Mpc$^{-1}$ reveals that the NNLO results of \gridspt\, covariance tend to slightly overpredict the $N$-body covariance, especially for the off-diagonal part of sphere 2. This is presumably because the survey window function of sphere 2 produces a rather tight correlation between large- and small-scale modes, and the off-diagonal covariance is largely affected by the small-scale modes for which the SPT predictions are no longer accurate, leading to a visible discrepancy. Nevertheless, the \gridspt\, covariance still provide an accurate quantitative description for the signal-to-noise ratio, as we shall see below.

%%--%%--%%--%%--%%--%%--%%--%%--%%--%%--%%--%%--%%--%%--%%--%%--%%--%%
\subsubsection{Signal-to-noise ratio}
\label{subsubsec:masked_covariance_snr}
%%--%%--%%--%%--%%--%%--%%--%%--%%--%%--%%--%%--%%--%%--%%--%%--%%--%%

Finally, using the full covariance matrix, we present the signal-to-noise ratio given at Eq.~(\ref{eq:def_snr}) in Fig.~\ref{fig:snr_gridSPT_z1_with_mask}.

In the presence of the survey masks, the signal-to-noise ratio in the Gaussian limit does not exactly follow the simple rule, $(S/N)= (k_{\rm max}V)^{3/2}/(12\pi^2)^{1/2}$. Here, the Gaussian results depicted as dotted lines are obtained from the linear-order \gridspt\, covariance, with the signal part (i.e., power spectrum) also evaluated with the linear power spectrum from \gridspt. Clearly, the achievable signal-to-noise ratios in the Gaussian limit depend on the survey masks, and among those we considered, the resultant $(S/N)$ for the survey mask of sphere 2 receives the largest impact.

The key finding here is that the \gridspt\, calculations accurately account for the survey window function effect and capture all the trends shown in $N$-body results. Similar to the case without survey mask, shown in Fig.~\ref{fig:snr_gridSPT_no_mask}, the signal-to-noise ratio estimated from the NLO results nicely agrees with that from the $N$-body simulations at $k\lesssim0.2\,h$\,Mpc$^{-1}$. Adding the NNLO, the agreement is further improved, and the estimated signal-to-noise ratios reproduce the simulations even at $k_{\rm max}\simeq0.35\,h$\,Mpc$^{-1}$.

%%%%%%%%%%%%%%%%%%%%%%%%%%%%%%%%%%%%%%%%%%%%%%%%%%%%%%%%%%%%%%%%%%%%%%%
%%%%%%%%%%%%%%%%%%%%%%%%%%%%%%%%%%%%%%%%%%%%%%%%%%%%%%%%%%%%%%%%%%%%%%%
\section{Discussions and conclusion}
\label{sec:conclusion}
%%%%%%%%%%%%%%%%%%%%%%%%%%%%%%%%%%%%%%%%%%%%%%%%%%%%%%%%%%%%%%%%%%%%%%%
%%%%%%%%%%%%%%%%%%%%%%%%%%%%%%%%%%%%%%%%%%%%%%%%%%%%%%%%%%%%%%%%%%%%%%%

In this paper, employing the perturbation theory to deal with the nonlinear evolution of large-scale structure, we have presented the accurate calculation of the power-spectrum covariance, taking also the effect of the survey window function into account. Our basis is a novel grid-based algorithm for the standard perturbation theory (SPT) calculations, which have been developed in Ref.~\cite{Taruya_Nishimichi_Jeong2018} and implemented in a {\tt c++} code, named \gridspt. 

Unlike the previous works using perturbation theory, our covariance calculations are not fully analytical, but rather numerical, similarly to those using $N$-body simulations. That is, we generate many realizations of the higher-order SPT density fields starting with random initial fields. Nevertheless, making use of the Fast-Fourier Transform, the \gridspt\, enables us to quickly generate those SPT density fields, which are then used to compute or measure the SPT power spectra at each perturbative order. We have given the recipes to reconstruct the power spectrum covariance perturbatively from the ensemble of SPT power spectra. The key expressions are given at Eqs.~(\ref{eq:PT_expansion_covariance})-(\ref{eq:PT_covariance_NNLO}). With an appropriate density estimator, these formulas can also be applied to the case including the survey window function and mask, and the covariance estimation can be made with \gridspt\, in a rather straightforward manner.

Our novel covariance calculation with \gridspt\, have been demonstrated both with and without incorporating the survey masks, respectively with the $100,000$ and $50,000$ realizations of the SPT density fields to fifth order in PT. The covariance matrices are then estimated including the non-Gaussian contributions arising from the non-vanishing trispectrum. The results containing the trispectrum at the tree-level (leading) and one-loop (next-to-leading) order are compared in detail with the measured covariance from $N$-body simulations. We find that that the \gridspt\, covariance at next-to-next-to-leading order (NLLO), which contains the one-loop trispectrum, quantitatively reproduces well the measured covariance in both cases with and without survey window function. A closer look at the small-scale behaviors reveals that the NNLO results of the \gridspt\, covariance tend to overpredict the simulations, especially for the off-diagonal part at the scales where the \gridspt\, fails to reproduce the power spectrum in $N$-body simulations. Still, the \gridspt\, covariance is shown to be useful in estimating the signal-to-noise ratio, and even on small scales, the NNLO covariance accurately explains the signal-to-noise ratio estimated from $N$-body simulations.

As discussed in Sec.~\ref{subsec:no_mask}, the single-stream PT calculation is known to be sensitively affected by the small-scale modes, and these UV-sensitive behaviors in SPT needs to be mitigated for a robust statistical predictions. To do so, the implementation of the effective-field-theory treatment (e.g., \cite{2012JCAP...07..051B,2012JHEP...09..082C,Nismichi_etal2020}) is important, and it would help improving the prediction. For future applications to observations, of crucial task is to incorporate the effects of redshift-space distortions and galaxy bias into \gridspt\, calculations. Also, the shot-noise contribution as well as the super-survey covariance are known to quantitatively give an impact on the covariance estimation \cite{Digvijay_Scoccimarro2019,Sugiyama_etal2019}, although the latter can be dealt with the so-called separate universe approach (e.g., \cite{Li_Hu_Takada2014a}), and hence can be easily implemented in the \gridspt\, calculation. Consistently incorporating all observational issues to the analytical PT calculation is rather challenging, but it is much simpler for \gridspt\, treatment. We will leave these issues for future work.

\acknowledgments
 We would like to thank Digvijay Wadekar for useful discussion. This work was supported in part by MEXT/JSPS KAKENHI Grant Number JP16H03977 and JP17H06359 (AT), and JP17K14273 and JP19H00677 (TN). AT and TN were also supported by JST AIP Acceleration Research Grant Number JP20317829, Japan. DJ was supported at Pennsylvania State University by NASA ATP program (80NSSC18K1103). Numerical computation was partly carried out at the Yukawa Institute Computer Facility. This research was also supported by the Munich Institute for Astro- and Particle Physics (MIAPP) which is funded by the Deutsche Forschungsgemeinschaft (DFG, German Research Foundation) under Germany's Excellence Strategy -- EXC-2094 -- 390783311.

\appendix

%%%%%%%%%%%%%%%%%%%%%%%%%%%%%%%%%%%%%%%%%%%%%%%%%%%%%%%%%%%%%%%%%%%%%%%
%%%%%%%%%%%%%%%%%%%%%%%%%%%%%%%%%%%%%%%%%%%%%%%%%%%%%%%%%%%%%%%%%%%%%%%
\section{Perturbative calculations of non-Gaussian covariance from \gridspt}
\label{sec:covariance_tree_1loop}
%%%%%%%%%%%%%%%%%%%%%%%%%%%%%%%%%%%%%%%%%%%%%%%%%%%%%%%%%%%%%%%%%%%%%%%
%%%%%%%%%%%%%%%%%%%%%%%%%%%%%%%%%%%%%%%%%%%%%%%%%%%%%%%%%%%%%%%%%%%%%%%

In this Appendix, the non-Gaussian covariance obtained from the \gridspt\, calculations is presented in the case without survey masks, particularly focusing on each building block of SPT calculations. 

Let us recall that in the absence of survey masks, the off-diagonal part of the covariance matrix purely represents the non-Gaussian contribution coming from the trispectrum [see Eq.~(\ref{eq:covariance})], and in the PT treatment of the covariance matrix, the off-diagonal part of the higher-order corrections, $\covSPTNLO$ and $\covSPTNNLO$, are respectively described by the tree-level and one-loop trispectra, $\overline{T}_{ij}^{\rm tree}$ and $\overline{T}_{ij}^{\rm 1\mbox{-}loop}$ [see Eqs.~(\ref{eq:covSPT_NLO_2nd}) and (\ref{eq:covSPT_NNLO_2nd})]. They are explicitly given by 
%%%%%%%%%%%%%%%%%%%%%%%%%%%%%%%%%%%%%%%%%%%%%%%%%%%%%%%%%%%%%%%%%%%%%%%
\begin{align}
\frac{\overline{T}_{ij}^{\rm tree}}{V}&= 
\Bigl\{\cov[\hat{P}_{11}(k_i),\hat{P}_{22}(k_j)] + 2\,\cov[\hat{P}_{11}(k_i),\hat{P}_{13}(k_j)] 
\nonumber
\\
&+(i\leftrightarrow j)\Bigr\} + 
 4\,\cov[\hat{P}_{12}(k_i),\hat{P}_{12}(k_j)],
\quad (i\ne j) 
\label{eq:trispectrum_tree}
\end{align}
%%%%%%%%%%%%%%%%%%%%%%%%%%%%%%%%%%%%%%%%%%%%%%%%%%%%%%%%%%%%%%%%%%%%%%%
and
%%%%%%%%%%%%%%%%%%%%%%%%%%%%%%%%%%%%%%%%%%%%%%%%%%%%%%%%%%%%%%%%%%%%%%%
\begin{align}
\frac{\overline{T}_{ij}^{\rm 1\mbox{-}loop}}{V} & =  
\Bigl\{2\,\cov[\hat{P}_{11}(k_i),\hat{P}_{15}(k_j)]
+2\, \cov[\hat{P}_{11}(k_i),\hat{P}_{24}(k_j)] 
\nonumber
\\
&+4\,\cov[\hat{P}_{12}(k_i),\hat{P}_{14}(k_j)] 
+ \cov[\hat{P}_{11}(k_i),\hat{P}_{33}(k_j)]
\nonumber
\\
&+2\,\cov[\hat{P}_{22}(k_i),\hat{P}_{13}(k_j)] 
+4\,\cov[\hat{P}_{12}(k_i),\hat{P}_{23}(k_j)]
\nonumber
\\
&+(i\leftrightarrow j)\Bigr\}
+4\,\cov[\hat{P}_{13}(k_i),\hat{P}_{13}(k_j)] 
\nonumber
\\
&+\cov[\hat{P}_{22}(k_i),\hat{P}_{22}(k_j)],
\quad (i\ne j).
\label{eq:trispectrum_1loop}
\end{align}
%%%%%%%%%%%%%%%%%%%%%%%%%%%%%%%%%%%%%%%%%%%%%%%%%%%%%%%%%%%%%%%%%%%%%%%
Below, based on the setup described in Sec.~\ref{subsec:setup}, the right-hand side of the expressions above is evaluated, and their off-diagonal parts at $z=0$ are separately plotted.

%%%%%%%%%%%%%%%%%%%%%%%%%%%%%%%%%%%%%%%%%%%%%%%%%%%%%%%%%%%%%%%%%%%%%%%
\begin{figure*}[tb]
\begin{center}
%\hspace*{0.5cm}
\includegraphics[height=7.4cm,angle=0]{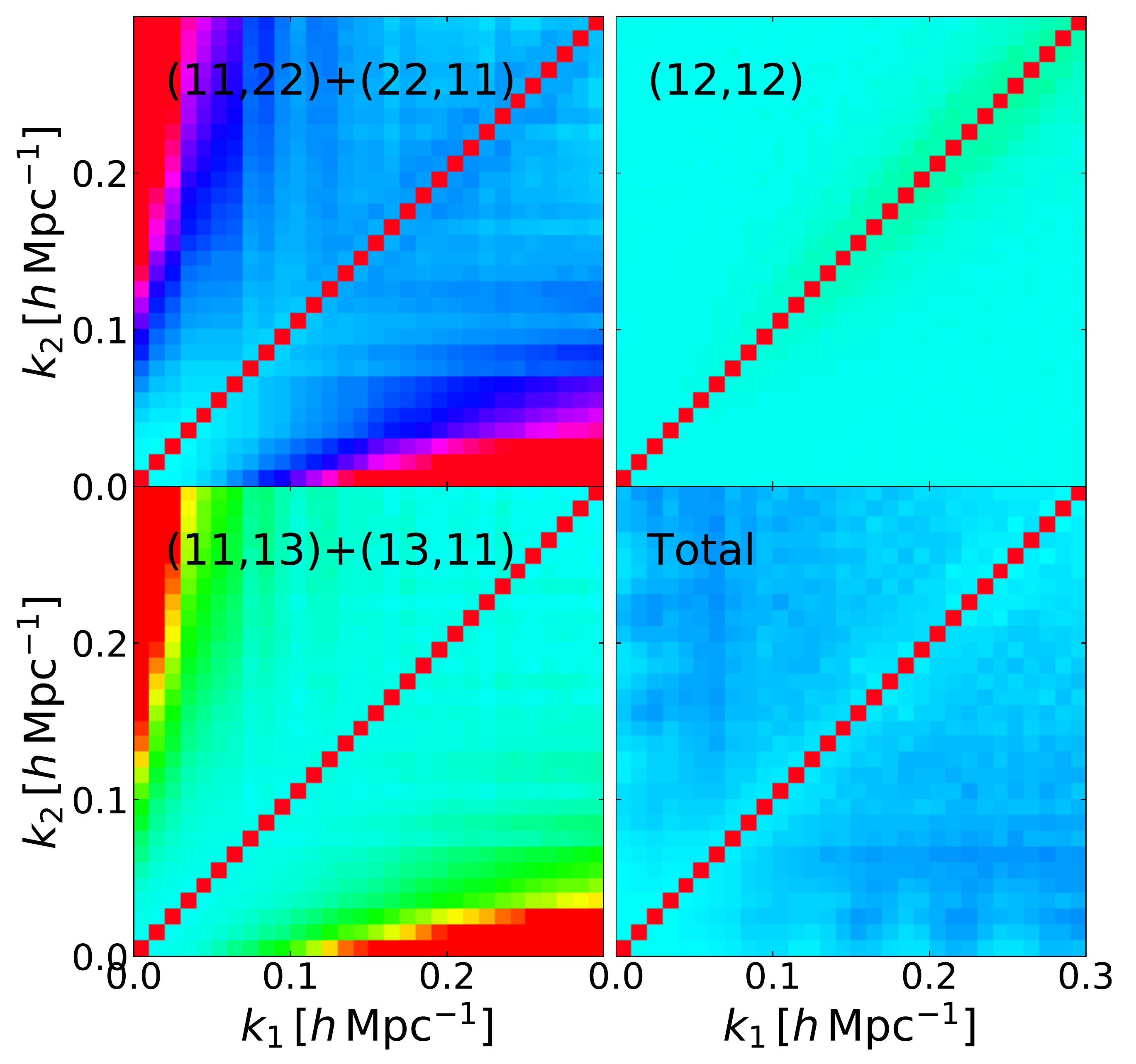}
%\hspace*{-0.8cm}
\includegraphics[height=7.4cm,angle=0]{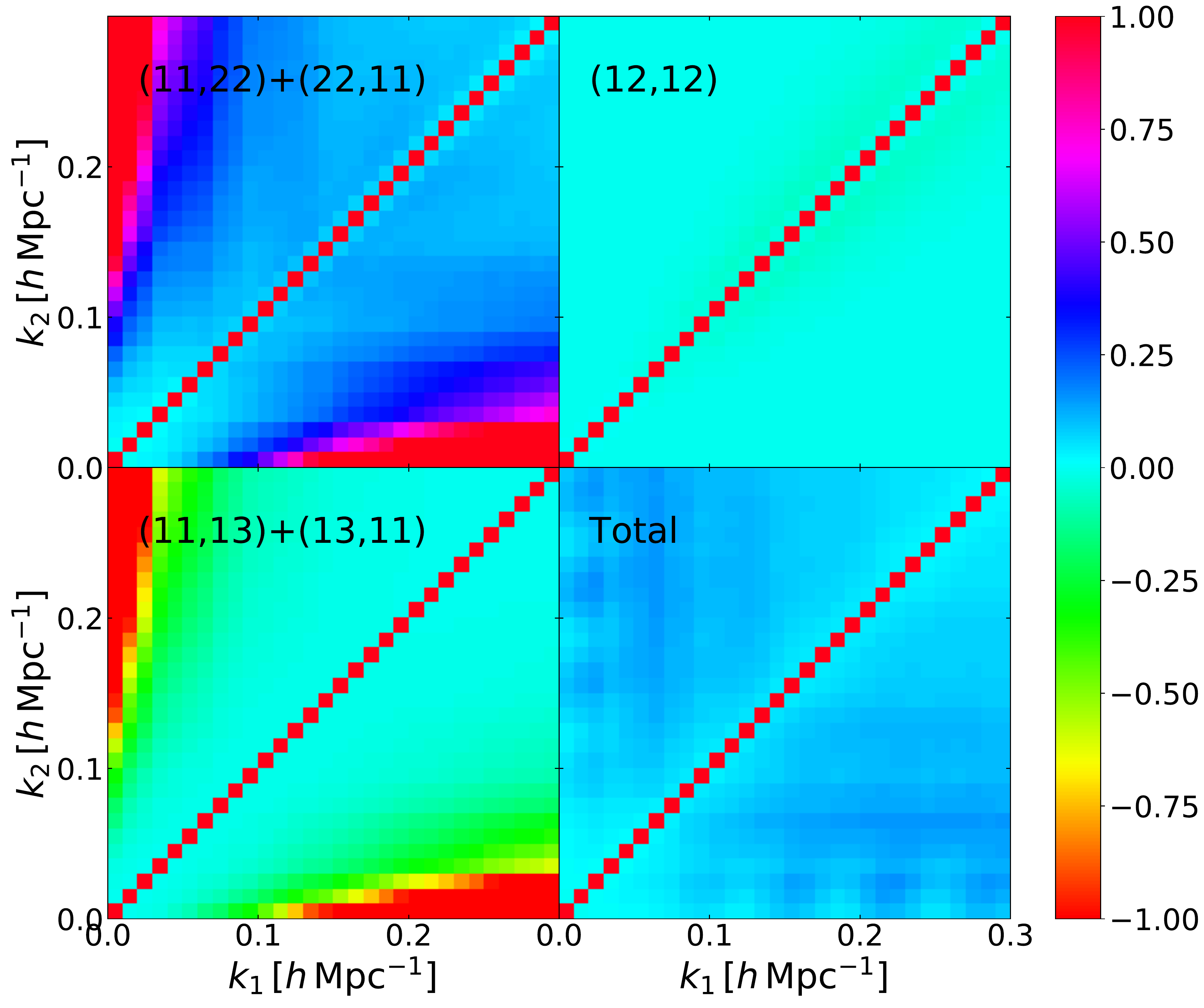}
\end{center}
\vspace*{-0.2cm}
\caption{Non-Gaussian covariance originated from the tree-level trispectrum. The results obtained from {\tt GridSPT} (left) and analytical SPT (right) calculations are shown. The plotted results are the correlation coefficient matrices $r(k_1,k_2)$ at $z=0$. Based on Eq.~(\ref{eq:trispectrum_tree}), we divide the trispectrum contribution into the three pieces, $(11,22)+(22,11)$, $(12,12)$, and $(11,13)+(13,11)$, which are separately shown, together with the sum of these contributions, $\overline{T}_{ij}^{\rm tree}/V$ (right bottom). Note that the diagonal components of each covariance matrix are replaced with those obtained from the $N$-body simulations. 
\label{fig:pkcov_tree_z0}
}
%\end{figure}
%%%%%%%%%%%%%%%%%%%%%%%%%%%%%%%%%%%%%%%%%%%%%%%%%%%%%%%%%%%%%%%%%%%%%%%
%%%%%%%%%%%%%%%%%%%%%%%%%%%%%%%%%%%%%%%%%%%%%%%%%%%%%%%%%%%%%%%%%%%%%%%
%\begin{figure}[tb]
%\vspace*{-0.8cm}
\begin{center}
%\hspace*{0.5cm}
\includegraphics[width=18cm,angle=0]{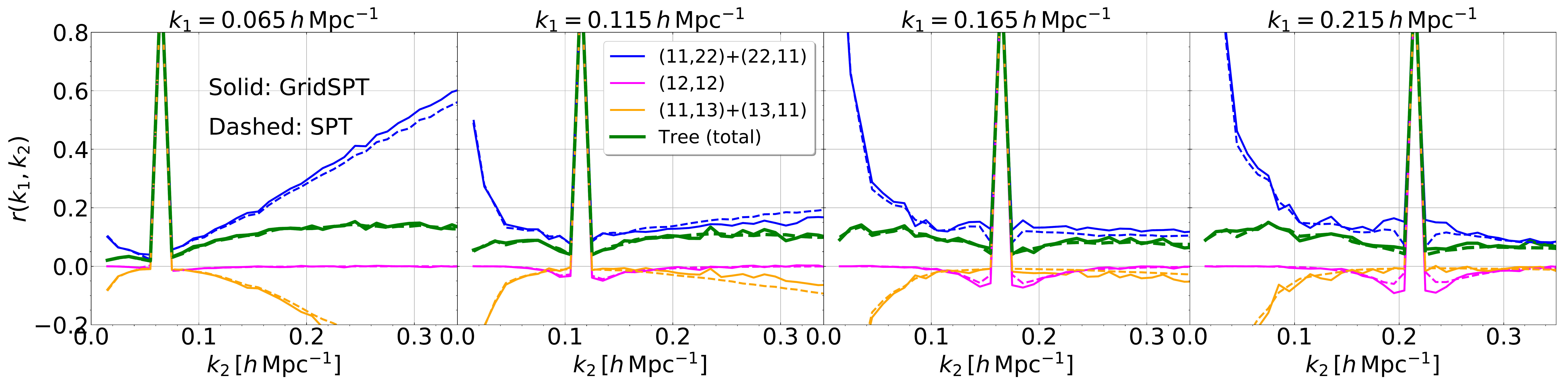}
\end{center}

\vspace*{-0.5cm}

\caption{Non-Gaussian covariance originated from the tree-level trispectrum, specifically fixing the wavenumber $k_1$ to $0.065$, $0.115$, $0.165$, and $0.215\,h$\,Mpc$^{-1}$ (from left to right). Similarly to Fig.~\ref{fig:pkcov_tree_z0}, the results obtained from \gridspt\,(solid) and analytical SPT (dashed) calculations are divided into three pieces, and for each, the correlation coefficient matrix is evaluated and plotted as a function of $k_2$. The green thick lines are the results summing up the three contributions. 
\label{fig:pkcov_tree_z0_k1_fixed}
}
\end{figure*}
%%%%%%%%%%%%%%%%%%%%%%%%%%%%%%%%%%%%%%%%%%%%%%%%%%%%%%%%%%%%%%%%%%%%%%%

Figs.~\ref{fig:pkcov_tree_z0} and \ref{fig:pkcov_tree_z0_k1_fixed} show the contributions from the tree-level trispectrum, which are divided into three pieces, $(11,22)+(22,11)$, $(12,12)$, and $(11,13)+(13,11)$. Here, the $(ab,cd)$ implies the covariance of the SPT power spectrum, $\cov[P_{ab}(k_1),P_{cd}(k_2)]$. Together with the total contribution, the correlation coefficient matrix $r(k_i,k_j)$ of each piece is compared with that of the analytical SPT results. Note that the color scale of Fig.~\ref{fig:pkcov_tree_z0} differs from those shown in Sec.~\ref{sec:results}. As we mentioned, the covariance estimated from \gridspt\, does not necessarily ensure the condition $|r(k_i,k_j)|\leq1$ [see Eq.~(\ref{eq:ratio_cov}) for definition], and some of the contributions eventually become larger (smaller) than $1$ $(-1)$ at $k_1\ll k_2$ and $k_1\gg k_2$. Nevertheless, summing up all the contributions, the cancellation happens, and the resultant value of the correlation coefficient matrix typically takes $r(k_i,k_j)\sim 0.1$ at $k_i\ne k_j$, as we have seen in Sec.~\ref{subsec:no_mask}. The estimated results of each contribution from \gridspt\, are all in good agreement with the analytical PT results.

%%%%%%%%%%%%%%%%%%%%%%%%%%%%%%%%%%%%%%%%%%%%%%%%%%%%%%%%%%%%%%%%%%%%%%%
\begin{figure*}[tb]
%\vspace*{-0.8cm}
\begin{center}
%\hspace*{0.5cm}
\includegraphics[width=12cm,angle=0]{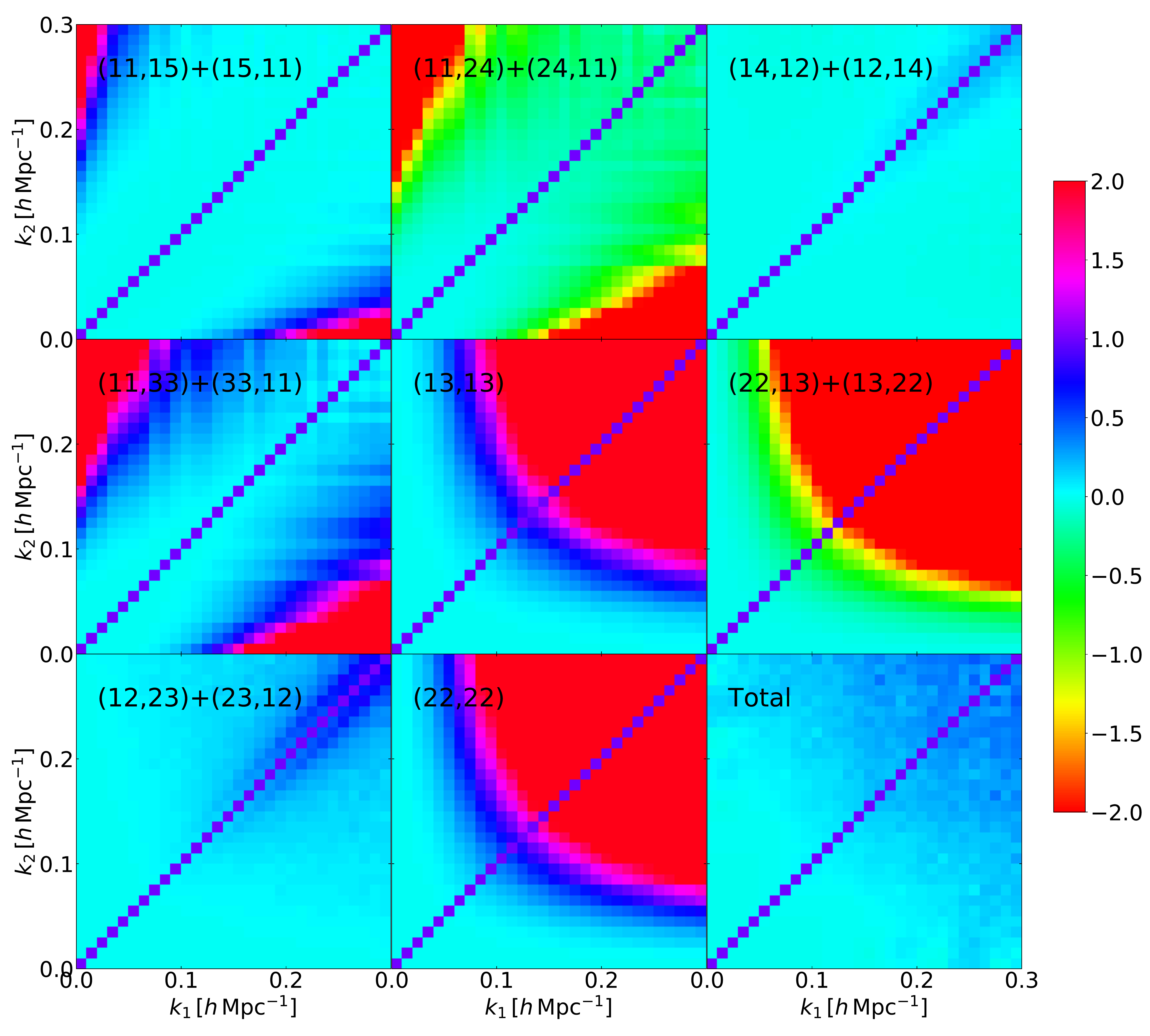}
\end{center}
\vspace*{-0.5cm}
\caption{Same as Fig.~\ref{fig:pkcov_tree_z0}, but the contributions originated from the one-loop trispectrum are shown, dividing the results into eight pieces according to Eq.~(\ref{eq:trispectrum_1loop}), i.e., $(11,15)+(15,11)$, $(11,24)+(24,11)$, $(12,14)+(14,12)$, $(11,33)+(33,11)$, $(13,13)$, $(22,13)+(13,22)$, $(12,23)+(23,12)$, $(22,22)$. The sum of these contributions, $\overline{T}_{ij}^{\rm 1\mbox{-}loop}/V$, or equivalently $\covSPTNNLO$, is also shown in right bottom panel. 
\label{fig:pkcov_1loop_z0}
}
%\end{figure}
%%%%%%%%%%%%%%%%%%%%%%%%%%%%%%%%%%%%%%%%%%%%%%%%%%%%%%%%%%%%%%%%%%%%%%%
%%%%%%%%%%%%%%%%%%%%%%%%%%%%%%%%%%%%%%%%%%%%%%%%%%%%%%%%%%%%%%%%%%%%%%%
%\begin{figure}[tb]
%\vspace*{-0.8cm}
\begin{center}
%\hspace*{0.5cm}
\includegraphics[width=18cm,angle=0]{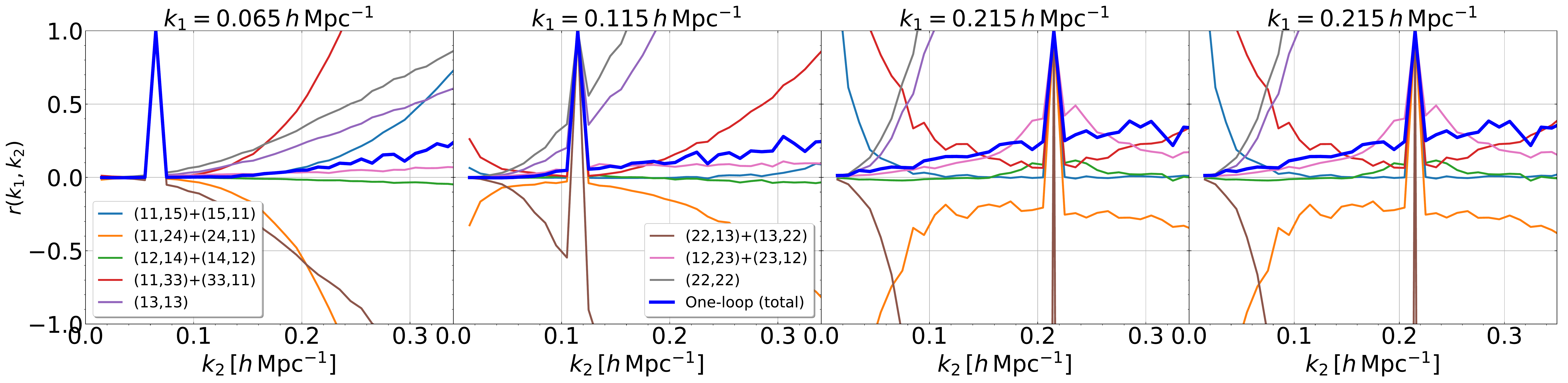}
\end{center}
\vspace*{-0.3cm}
\caption{Same as Fig.~\ref{fig:pkcov_tree_z0_k1_fixed}, but the contributions originated from the one-loop trispectrum are shown, dividing the results into eight pieces, similarly to Fig.~\ref{fig:pkcov_1loop_z0}. The blue thick lines are the results summing up all contributions, which correspond to $\covSPTNNLO$. 
\label{fig:pkcov_1loop_z0_k1_fixed}
}
\end{figure*}
%%%%%%%%%%%%%%%%%%%%%%%%%%%%%%%%%%%%%%%%%%%%%%%%%%%%%%%%%%%%%%%%%%%%%%%
Next look at the contributions coming from the one-loop trispectrum. Here, dividing the non-Gaussian contributions into eight pieces, their results at $z=0$ are plotted in Figs.~\ref{fig:pkcov_1loop_z0} and \ref{fig:pkcov_1loop_z0_k1_fixed}, together with the sum of these results. Again, the color scales of Figs.~\ref{fig:pkcov_1loop_z0} and plot range of vertical axis in Fig.~\ref{fig:pkcov_1loop_z0_k1_fixed} have been changed. Similarly to the previous case, we see a rather large change in $r(k_1,k_2)$ not only at the region of $k_1\ll k_2\,(k_2\gg k_1)$ but also at $k_{1,2} \gtrsim0.1\,h$\,Mpc$^{-1}$. The amplitude of each contribution gets also larger, with either positive or negative sign. However, the cancellation again happens, and the sum of all the one-loop corrections approaches zero at large scales, known as a consequence of the Galilean invariance in SPT calculations.

%%%%%%%%%%%%%%%%%%%%%%%%%%%%%%%%%%%%%%%%%%%%%%%%%%%
%%%%%%%%%%%%%%%%%%%%%%%%%%%%%%%%%%%%%%%%%%%%%%%%%%%
%%%%%%%%%%%%%%%%%%%%%%%%%%%%%%%%%%%%%%%%%%%%%%%%%%%%%%%
%\bibliography{ref}
\bibliographystyle{apsrev4-1}
\input{references.bbl}
%%%%%%%%%%%%%%%%%%%%%%%%%%%%%%%%%%%%%%%%%%%%%%%%%%%%%%%
%%%%%%%%%%%%%%%%%%%%%%%%%%%%%%%%%%%%%%%%%%%%%%%%%%%

\end{document}